\documentstyle[11pt,axodraw,ijmpa1]{article}
\begin{document}
\setcounter{footnote}{0}
\renewcommand{\thefootnote}{\alph{footnote}}
\renewcommand{\theequation}{\thesection.\arabic{equation}}
\newcounter{saveeqn}
\newcommand{\add}{\addtocounter{equation}{1}}
\newcommand{\alpheqn}{\setcounter{saveeqn}{\value{equation}}%
\setcounter{equation}{0}%
\renewcommand{\theequation}{\mbox{\thesection.\arabic{saveeqn}{\alph{equation}}}}}
\newcommand{\reseteqn}{\setcounter{equation}{\value{saveeqn}}%
\renewcommand{\theequation}{\thesection.\arabic{equation}}}
\newenvironment{nedalph}{\add\alpheqn\begin{eqnarray}}{\end{eqnarray}\reseteqn}
\newsavebox{\PSLASH}
\sbox{\PSLASH}{$p$\hspace{-1.8mm}/}
\newcommand{\PS}{\usebox{\PSLASH}}
\newsavebox{\PARTIALSLASH}
\sbox{\PARTIALSLASH}{$\partial$\hspace{-2.3mm}/}
\newcommand{\PARTIALS}{\usebox{\PARTIALSLASH}}
\newsavebox{\ASLASH}
\sbox{\ASLASH}{$A$\hspace{-2.1mm}/}
\newcommand{\AS}{\usebox{\ASLASH}}
\newsavebox{\KSLASH}
\sbox{\KSLASH}{$k$\hspace{-1.8mm}/}
\newcommand{\KS}{\usebox{\KSLASH}}
\newsavebox{\LSLASH}
\sbox{\LSLASH}{$\ell$\hspace{-1.8mm}/}
\newcommand{\LS}{\usebox{\LSLASH}}
\newsavebox{\SSLASH}
\sbox{\SSLASH}{$s$\hspace{-1.8mm}/}
\newcommand{\SS}{\usebox{\SSLASH}}
\newsavebox{\DSLASH}
\sbox{\DSLASH}{$D$\hspace{-2.8mm}/}
\newcommand{\DS}{\usebox{\DSLASH}}
\newsavebox{\DbfSLASH}
\sbox{\DbfSLASH}{${\mathbf D}$\hspace{-2.8mm}/}
\newcommand{\DBFS}{\usebox{\DbfSLASH}}
\newsavebox{\DELVECRIGHT}
\sbox{\DELVECRIGHT}{$\stackrel{\rightarrow}{\partial}$}
\newcommand{\PARVECR}{\usebox{\DELVECRIGHT}}
\thispagestyle{empty}
\begin{flushright}
IPM/P-2000/045
\par
hep-th/0009233
\end{flushright}
\vspace{0.5cm}
\begin{center}
{\Large\bf{Anomaly and Nonplanar Diagrams in Noncommutative Gauge Theories}}\\
\vspace{1cm}
{\bf Farhad Ardalan}\footnote{\normalsize{Electronic address: ardalan@theory.ipm.ac.ir}}
\hspace{0.2cm}
and \hspace{0.2cm}{\bf N\'eda Sadooghi}\footnote{\normalsize{Electronic address:
sadooghi@theory.ipm.ac.ir}}  \\
\vspace{0.5cm}
{\sl Institute for Studies in Theoretical Physics and Mathematics (IPM)}\\
{\sl{School of Physics, P.O. Box 19395-5531, Tehran-Iran}}\\
and\\
{\sl Department of Physics, Sharif University of Technology}\\
{\sl P.O. Box 11365-9161, Tehran-Iran}
\end{center}
\vspace{0cm}
\begin{center}
{\bf {Abstract}}
\end{center}
\begin{quote}
Anomalies arising from nonplanar triangle diagrams of noncommutative gauge theory are studied.
Local chiral gauge anomalies for both noncommutative U(1) and U(N) gauge theories with adjoint
matter fields are shown to vanish. For noncommutative QED with fundamental matters, due to UV/IR mixing a finite anomaly emerges from the nonplanar contributions. It involves a generalized $\star$-product of gauge fields.   
\end{quote}

\hspace{0.8cm}
\par\noindent
{\it PACS No.:} 11.15.Bt, 11.10.Gh, 11.25.Db
\par\noindent
{\it Keywords:} Noncommutative Field Theory, Chiral Gauge Anomaly, Global Anomaly 
\newpage
\setcounter{page}{1}
\section{Introduction}
\setcounter{section}{1}
\noindent
Noncommutative geometry\cite{connes} has made a dramatic appearance in string theory
recently\cite{witten} and has made noncommutative gauge theory\cite{connes} an active
field of study. The various aspects of noncommutative Yang-Mills (NCYM) theory have
been extensively studied\cite{ym,rest} and a number of novel phenomena
discovered\cite{open,solitons,minwalla,uvir}. Of special importance is a certain 
duality between ultraviolet (UV) and infrared (IR) behavior of nonplanar loop
diagrams of NC-field theories and in particular NCYM theories, 
which manifests itself in the singularity of amplitudes in two limits of 
small noncommutativity parameter $\theta$ and large momentum 
cutoff $\Lambda$ of the theory\cite{minwalla}.
\par
Recently Seiberg\cite{seiberg} has observed that NCYM theory 
appears as a manifestation of a matrix model and the underlying model forces the coupling 
of the matter to gauge fields to be in the adjoint representation of the noncommutative algebra.
But then, it is known that only in the adjoint coupling there appear nonplanar diagrams
 and there is  a chance of observing the UV/IR mixing\cite{uvir}. 
\par
Anomalies of NCYM have been studied in a number of papers\cite{neda3,chiral} and it has
 been found that for the fundamental and anti-fundamental fermion coupling, replacing the
 usual products by $\star$-products will give the anomaly modulo certain subtlties\cite{neda3}.
 This is because only planar diagrams appear in the triangle anomaly for this coupling.
 This result was also confirmed in the Fujikawa path integral method\cite{neda3,chiral}.
\par
In this paper we study the effect of nonplanar diagrams on anomalies. We will find that
 nonplanar diagrams do not contribute to gauge anomalies. However, global symmetries reflect
 the unconventional behavior of the nonplanar diagrams in the form of a contribution to the
 global anomaly which involves a generalized $\star$-product. 
\par\noindent
In the first part of the paper, we will discuss the anomalies in gauge symmetry of the
 noncommutative U(1) and U(N) theories. Since we are interested in the effect of nonplanar
 integrals on the anomaly, the matter fields in both theories are taken to be in the adjoint
 representation. We will first calculate the chiral anomaly for U(1) theory where  planar as well as nonplanar diagrams will appear. 
Both contributions to the triangle anomaly vanish. The chiral anomaly of U(N) theory with
 matter fields in the adjoint representation of both gauge group and the noncommutative
 algebra, is then calculated. Here also, planar as well as nonplanar contributions  vanish. Hence both 
theories are free of gauge anomaly\footnote{When this study was almost done an article by C. P. Martin\cite{martin} appeared, with the same result for U(N) theory calculated in a different manner.}. In the ordinary commutative field theory no anomaly will appear when the group representation is real. In noncommutative the concept of a real representation does not obviously apply, since {\it both} gauge group and matter field coupling are in the adjoint representation.   
\par 
In the second part of the paper, we discuss the global symmetry of the U(1) theory with matter fields in fundamental representation. There are three different currents, corresponding to three different local change of variables. We will discuss their global and local properties in two different cases of spacelike and timelike or lightlike noncommutativity. In a noncommutative theory containing fundamental matter fields,  the triangle diagrams for the first current include only planar contributions and leads to the usual anomaly of the  noncommutative U(1), calculated previously\cite{neda3}. The second current exhibits, due to the UV/IR mixing\cite{minwalla}, a finite anomaly arising only from nonplanar diagrams. The result of the anomaly contains an unusual form of $\star$-product. These generalized $\star$-products appeared recently in some other works\cite{garousi,liustar,reststar,ken}.  
\par\vskip0.5cm
\section{Chiral Gauge Anomaly, Adjoint Matter Fields }
\par\vskip0.2cm\noindent
{\it i) Noncommutative U(1)}
\setcounter{section}{2}
\setcounter{equation}{0}
\par\vskip0.2cm\noindent
We follow the notation of Ref.\cite{neda3} and recall that the noncommutative (NC) gauge theory is characterized by the replacement of the familiar product of functions with the $\star$-product defined by:
\begin{eqnarray}\label{M2-1-1}  
f\left(x\right)\star g\left(x\right)\equiv e^{\frac{i\theta_{\mu\nu}}{2}\ \frac{\partial}{\partial\xi_{\mu}}\ \frac{\partial}{\partial\zeta_{\nu}} }f\left(x+\xi\right)g\left(x+\zeta\right)\bigg|_{\xi=\zeta=0},
\end{eqnarray}
where $\theta_{\mu\nu}$ is a real constant antisymmetric back\-ground, and re\-flects the noncom\-mu\-ta\-ti\-vi\-ty of the coordinates
\begin{eqnarray}\label{M2-1-2}
[x_{\mu},x_{\nu}]=i\theta_{\mu\nu}.
\end{eqnarray} 
In NC-U(1) gauge theory the matter fields can couple to gauge fields in fundamental,  anti-fun\-da\-men\-tal, or adjoint representation. The fundamental and anti-fundamental covariant derivatives are given by
\begin{eqnarray}\label{M2-1-3}
\hspace{-0.5cm}D_{\mu}\psi_{L}\left(x\right)\equiv \partial_{\mu}\psi_{L}\left(x\right)+ig\ A_{\mu}\left(x\right)\star \psi_{L}\left(x\right),&\mbox{and}&D_{\mu}\psi\left(x\right)\equiv\partial_{\mu}\psi_{L}\left(x\right)-ig \psi_{L}\left(x\right)\star A_{\mu}\left(x\right),
\end{eqnarray}
respectively. In the adjoint representation, it is defined by
\begin{eqnarray}\label{M2-1-4}
D_{\mu}\psi_{L}\left(x\right)\equiv \partial_{\mu}\psi_{L}\left(x\right)+ig[A_{\mu}\left(x\right),\psi_{L}\left(x\right)]_{\star}.
\end{eqnarray}
Here, the left-handed fermions are given by $\psi_{L}\equiv P_{+}\psi$ with $P_{+}\equiv
 \frac{1}{2}\left(1+\gamma_{5}\right)$ and $\overline{\psi}_{L}\equiv \overline{\psi}P_{-}$
 with $P_{-}\equiv \frac{1}{2}\left(1-\gamma_{5}\right)$. 
Using these notations the corresponding fermionic action for NC-U(1) with matter field in 
the adjoint representation is given by:
\begin{eqnarray}\label{M2-1-5}
S_{F}[\overline{\psi},\psi]=\int d^{D}x \bigg\{i\overline{\psi}_{\alpha}\left(x\right)\star\partial_{\mu}\psi_{\beta}\left(x\right)-g\overline{\psi}_{\alpha}\left(x\right)\star \big[A_{\mu}\left(x\right),\psi_{\beta}\left(x\right)\big]_{\star}\bigg\}\left(\gamma_{\mu}P_{+}\right)^{\alpha\beta}.
\end{eqnarray}
This action is invariant under the following local $\star$-gauge transformations
\begin{eqnarray}\label{M2-1-6}
\overline{\psi}\left(x\right)\rightarrow \overline{\psi}'\left(x\right)=
U\left(x\right)\star\overline{\psi}\left(x\right)\star U^{-1}\left(x\right),&\hspace{-0.4cm}&\psi\left(x\right)\rightarrow \psi'\left(x\right)=
U\left(x\right)\star\psi\left(x\right)\star U^{-1}\left(x\right),
\end{eqnarray}
and 
\begin{eqnarray}\label{M2-1-7}
A'_{\mu}\left(x\right)=U\left(x\right)\star A_{\mu}\left(x\right)\star U^{-1}\left(x\right)-\frac{i}{g}U\left(x\right)\star\partial_{\mu}U^{-1}\left(x\right).
\end{eqnarray}
Here, $U\left(x\right)\equiv \left(e^{ig\alpha\left(x\right)}\right)_{\star}$ and $\alpha\left(x\right)$ is an arbitrary function. The local current corresponding to the above action is given by:
\begin{eqnarray}\label{M2-1-8}
J_{\mu}\left(x\right)\equiv -i\left(\gamma_{\mu}P_{+}\right)^{\alpha\beta} \bigg\{\overline{\psi}_{\alpha}\left(x\right),\psi_{\beta}\left(x\right)\bigg\}_{\star},
\end{eqnarray}
where $\{\overline{\psi}_{\alpha}\left(x\right),\psi_{\beta}\left(x\right)\}_{\star}\equiv \overline{\psi}_{\alpha}\left(x\right)\star\psi_{\beta}\left(x\right)+\psi_{\beta}\left(x\right)\star\overline{\psi}_{\alpha}\left(x\right)$. 
\par
To study the gauge invariance and to determine the chiral anomaly  we consider the three-point function
\begin{eqnarray}\label{M2-1-9}
\Gamma_{\mu\lambda\nu}\left(x,y,z\right)\equiv\bigg<T\left(
J_{\mu}\left(x\right)J_{\lambda}\left(y\right)J_{\nu}\left(z\right)\right)
\bigg>.
\end{eqnarray}
Contracting the fermionic fields gives two types of triangle diagrams [See Fig. \ref{Fig1}]. The corresponding Feynman integrals are given by: 
\begin{eqnarray}\label{M2-1-10}
\lefteqn{\hspace{-0.8cm}
\Gamma_{\mu\lambda\nu}\left(x,y,z\right)=
-\int\limits_{-\infty}^{+\infty}
\frac{d^{4}k_{2}}{\left(2\pi\right)^{4}}
\frac{d^{4}k_{3}}{\left(2\pi\right)^{4}}
\frac{d^{4}\ell}{\left(2\pi\right)^{4}}
e^{-i\left(k_{2}+k_{3}\right)x}\ e^{ik_{2}y}e^{ik_{3}z}
}\nonumber\\
&&\hspace{-1.cm} \times \Bigg\{
\bigg[\mbox{Tr}\left(D^{-1}\left(\ell-k_{3}\right)\gamma_{\mu}P_{+}D^{-1}\left(\ell+k_{2}\right)\gamma_{\lambda} D^{-1}\left(\ell\right)
\gamma_{\nu}\right)\bigg]\ F_{a}\left(\ell\right)+
\big[\left(k_{2},\lambda\right)\leftrightarrow\left(k_{3},\nu\right)\big]
\ F_{b}\left(\ell\right)\Bigg\},\nonumber\\
\end{eqnarray}
where for massless fermions $D\left(\ell\right)\equiv \LS$. The phase factors are
\begin{nedalph}\label{M2-1-11a}
F_{a}\left(\ell\right)&=&e^{i\ k_{2}\times k_{3}}\bigg[
1- e^{2i\ell\times k_{3} }
-e^{2i\ell\times k_{2} }
+e^{2i\ell\times \left(k_{3}+k_{2}\right) }
\bigg]\nonumber\\
&&\hspace{-0.5cm}
+e^{-i\ k_{2}\times k_{3}}\bigg[-1+
e^{-2i\ell\times k_{3} }
+e^{-2i\ell\times k_{2} }
-e^{-2i\ell\times \left(k_{2}+k_{3}\right) }
\bigg],
\end{eqnarray}
\begin{eqnarray}\label{M2-1-11b}
\hspace{-7cm}\mbox{and}\hspace{4cm}F_{b}\left(\ell\right)=-F_{a}\left(\ell\to -\ell\right).
\end{nedalph}
Here $p\times q\equiv \frac{1}{2}\theta_{\eta\sigma}p^{\eta}q^{\sigma}$. Going through standard algebraic manipulations shows that the local gauge invariance is intact despite the appearance of nonplanar integrals. 
\par
To determine the chiral anomaly of $\partial^{\mu}\Gamma_{\mu\lambda\nu}$, we use dimensional regularization. Recall that in the dimensional regularization, 
$\gamma_{5}$ is defined so that it anticommutes with $\gamma_{\mu}$'s for $\mu=0,1,2,3$ but commutes with them for other values of $\mu$, and that the loop momenta have components in all dimensions, whereas the external momenta $k_{2}$ and $k_{3}$ are only four dimensional. Using $\ell=\ell_{||}+\ell_{\perp}$, where $\ell_{||}$ has nonzero components in dimensions $0,1,2,3$ and $\ell_{\perp}$ has nonzero components in the other $D-4$ dimensions, and the identity 
\begin{nedalph}\label{M2-1-16a}
\left(\KS_{2}+\KS_{3}\right)P_{+}=P_{-}D\left(\ell+k_{2}\right)-D\left(\ell-k_{3}\right)P_{+}+\gamma_{5}\LS_{\perp},
\end{eqnarray}
in the corresponding integrals of diagram $A$ and
\begin{eqnarray}\label{M2-1-16b}
\left(\KS_{2}+\KS_{3}\right)P_{+}=P_{-}D\left(\ell+k_{3}\right)-D\left(\ell-k_{2}\right)P_{+}+\gamma_{5}\LS_{\perp}.
\end{nedalph}
in the Feynman integral corresponding to the diagram $B$, we arrive at:
\begin{nedalph}\label{M2-1-17a}
\partial^{\mu}\Gamma_{\mu\lambda\nu}=
-i\int\limits_{-\infty}^{+\infty}
\frac{d^{4}k_{2}}{\left(2\pi\right)^{4}}
\frac{d^{4}k_{3}}{\left(2\pi\right)^{4}}\ e^{-i\left(k_{2}+k_{3}\right) x}
\ e^{ik_{2}y}\ e^{ik_{3}z}
 \bigg[A_{\lambda\nu}\left(k_{2},k_{3};\theta\right)+R_{\lambda\nu}\left(k_{2},k_{3};\theta\right)\bigg],
\end{eqnarray}
where
\begin{eqnarray}\label{M2-1-17b}
A_{\lambda\nu}\left(k_{2},k_{3};\theta\right)
&\equiv & \int\limits_{-\infty}^{+\infty}\frac{d^{D}\ell}{\left(2\pi\right)^{D}}
\Bigg[\mbox{Tr}\left( D^{-1}\left(\ell-k_{3}\right)\gamma_{5}\LS_{\perp}D^{-1}\left(\ell+k_{2}\right)\gamma_{\lambda}D^{-1}\left(\ell\right)\gamma_{\nu}\right)\ F_{a}\left(\ell\right)\Bigg]\nonumber\\
&&+\big[\left(k_{2},\lambda\right)\leftrightarrow \left(k_{3},\nu\right)\big]F_{b}\left(\ell\right),
\end{eqnarray}
and 
\begin{eqnarray}\label{M2-1-17c}
\lefteqn{R_{\lambda\nu}\left(k_{2},k_{3};\theta\right)\equiv}\nonumber\\
&&\hspace{-0.6cm}\equiv\int\limits_{-\infty}^{+\infty}\frac{d^{D}\ell}{\left(2\pi\right)^{D}}
\Bigg\{\bigg[
\mbox{Tr}\left(D^{-1}\left(\ell-k_{3}\right)P_{-}\gamma_{\lambda}D^{-1}\left(\ell\right)\gamma_{\nu}\right)-
\mbox{Tr}\left(P_{+}D^{-1}\left(\ell+k_{2}\right)\gamma_{\lambda}D^{-1}\left(\ell\right)\gamma_{\nu}\right)\bigg]
F_{a}\left(\ell\right)\nonumber\\
&&\hspace{-0.5cm}+\bigg[
\mbox{Tr}\left(D^{-1}\left(\ell-k_{2}\right)P_{-}\gamma_{\nu}D^{-1}\left(\ell\right)\gamma_{\lambda}\right)-
\mbox{Tr}\left(P_{+}D^{-1}\left(\ell+k_{3}\right)\gamma_{\nu}D^{-1}\left(\ell\right)\gamma_{\lambda}\right)
\bigg]F_{b}\left(\ell\right)\Bigg\}.
\end{nedalph}
The contributions of the planar phases to $A_{\lambda\nu}$ from Eq. (\ref{M2-1-17b}) vanish, because as we know from ordinary commutative U(1) the contributions of both relevant triangle diagrams are equal due to Bose symmetry. In the noncommutative U(1), the planar part of $A_{\lambda\nu}$ are just the same as their commutative counterparts, except here these integrals are to be modified with corresponding planar phase factors, which can be taken out of the integration. Hence in order to add the contributions of diagrams $A$ and $B$ to $A_{\lambda\nu}^{pl.}$ only the planar phases have to be added together. The planar phase for the diagram $A$ turns out to be $2i\sin\left( k_{2}\times k_{3}\right)$ whereas for the diagram $B$ is $-2i\sin\left( k_{2}\times k_{3}\right)$ [see Eqs. (\ref{M2-1-11a}) and (\ref{M2-1-11b})]. Hence the planar contribution of $B$ cancels the planar contribution of $A$. 
\par
For the nonplanar part the above argument does not go through and the calculation must be performed explicitly. After appropriate shift of integration variables and using (\ref{M2-1-11b}) for nonplanar phases, it can be shown that the contribution of diagram A cancels the contribution of diagram B, so that the nonplanar part of $A_{\lambda\nu}$ vanishes too.  Hence, $\partial^{\mu}\Gamma_{\mu\lambda\nu}$ does not receive any anomalous contribution  from $A_{\lambda\nu}$ [Eq. (\ref{M2-1-17b})]. 
\par
The contribution of $R_{\lambda\nu}$ can also be shown not to contribute to anomaly. 
Noncommutative U(1) with adjoint matter fields is therefore free of chiral gauge anomaly. 
\par\vskip0.5cm\noindent
{\it ii) Noncommutative U(N)}
\par\vskip0.2cm\noindent
Let us introduce the matter fields in NC-U(N) with the covariant derivative:
\begin{eqnarray}\label{M2-2-1}
{\mathbf D}_{\mu}\Psi_{L}\left(x\right)\equiv \partial_{\mu}\Psi_{L}\left(x\right)+ig[{\mathbf A}_{\mu}\left(x\right),\Psi_{L}\left(x\right)]_{\star},
\end{eqnarray} 
where ${\mathbf A}_{\mu}\equiv A_{\mu}^{a}t^{a}$, $\Psi_{L}\equiv \psi^{a}_{L}t^{a}$ and $t^{a}$, $a=0,\cdots N^{2}-1$, are the generators of U(N) and $\psi^{a}_{L}\equiv P_{+}\psi^{a}$ and $P_{+}=\frac{1}{2}\left(1+\gamma_{5}\right)$. It is given explicitly by
\begin{eqnarray}\label{M2-2-2}
{\mathbf D}_{\mu}\psi^{a}_{L}\equiv \partial_{\mu}\psi^{a}_{L}+\frac{ig}{2}D^{abc}[A_{\mu}^{b},\psi^{c}_{L}]_{\star}-\frac{g}{2}C^{abc}\{A_{\mu}^{b},\psi^{c}_{L}\}_{\star}.
\end{eqnarray}
Here, we have used the identity $2t^{a}t^{b}=D^{abc}t^{c}+iC^{abc}t^{c}$,
where the $D^{abc}$ and $C^{abc}$ are given by  $\{t^{a},t^{b}\}\equiv D^{abc}t^{c}$ and $[t^{a},t^{b}]\equiv iC^{abc}t^{c}$. Using the definitions (\ref{M2-2-1}) and (\ref{M2-2-2}) of the covariant derivative, the fermionic action of this model is given by:
\begin{eqnarray}\label{M2-2-4}
\lefteqn{
{\cal{S}}\big[\overline{\psi}^{a},\psi^{a},A_{\mu}^{a} \big]=\int\limits_{-\infty}^{+\infty}d^{4}x\  \left(\gamma^{\mu}P_{+}\right)_{\alpha\beta}{\cal{N}}\Bigg[i\left(\overline{\psi}^{a}_{\alpha}\left(x\right)\star\partial_{\mu}\psi^{a}_{\beta}\left(x\right)\right)
}\nonumber\\
&&-\frac{g}{2} \left(C^{abc}\overline{\psi}^{a}_{\alpha}\left(x\right)\star\{A_{\mu}^{b}\left(x\right),\psi^{c}_{\beta}\left(x\right)\}_{\star}-iD^{abc}\overline{\psi}^{a}_{\alpha}\left(x\right)\star \big[A_{\mu}^{b}\left(x\right),\psi^{c}_{\beta}\left(x\right)\big]_{\star}\right)\bigg],
\end{eqnarray}
where Tr$\left(t^{a}t^{b}\right)={\cal{N}}\delta^{ab}$ is used. This action is invariant under infinitesimal gauge transformations
\begin{nedalph}\label{M2-2-5a}
\delta_{\alpha}\overline{\Psi}=-ig\big[\overline{\Psi}\left(x\right),\alpha\left(x\right)\big]_{\star},\hspace{0.5cm}\mbox{and}\hspace{0.5cm}
\delta_{\alpha}\Psi=+ig\big[\alpha\left(x\right),\Psi\left(x\right)\big]_{\star},
\end{eqnarray}
and 
\begin{eqnarray}\label{M2-2-6a}
\delta_{\alpha}{\mathbf A}_{\mu}(x)=-\partial_{\mu}\alpha(x)+ig\big[\alpha\left(x\right),{\mathbf A}_{\mu}\left(x\right)\big]_{\star}.
\end{nedalph}
This can be established after a lengthy but straightforward calculation using the Jacobi identities given in the Eq. (\ref{A1}) [See appendix A]. In this model the local current reads
\begin{nedalph}\label{M2-2-7a}
{\mathbf J}_{\mu}\left(x\right)\equiv -\frac{i}{{\cal{N}}}\bigg\{\Psi_{L}^{\alpha}\left(x\right),\overline{\Psi}_{L}^{\beta}\left(x\right)\bigg\}_{\star}\left(\gamma_{\mu}\right)_{\alpha\beta}, 
\end{eqnarray} 
where ${\mathbf J}_{\mu}=J^{a}_{\mu} t^{a}$. The components of the current read
\begin{eqnarray}\label{M2-2-7b}
J^{a}_{\mu}\left(x\right)=\frac{-i}{2}\left(D^{abc}\big\{ \psi_{\alpha}^{b}\left(x\right),\overline{\psi}_{\xi}^{c}\left(x\right) \big\}_{\star}+iC^{abc}\big[\psi_{\alpha}^{b}\left(x\right),\overline{\psi}_{\xi}^{c}\left(x\right) \big]_{\star}\right)\left(\gamma_{\mu}P_{+}\right)^{\xi\alpha}.
\end{nedalph}
To keep the calculation as short as possible we have used the following form of $J_{\mu}^{a}\left(x\right)$:
\begin{eqnarray}\label{M2-2-8}
J^{a}_{\mu}\left(x\right)=-i\left(K^{abc}\big\{ \psi_{\alpha}^{b}\left(x\right),\overline{\psi}_{\xi}^{c}\left(x\right) \big\}_{\star}-iC^{abc}\overline{\psi}_{\xi}^{c}\left(x\right)\star\psi_{\alpha}^{b}\left(x\right)\right)\left(\gamma_{\mu}P_{+}\right)^{\xi\alpha},
\end{eqnarray} 
where $K^{abc}=\frac{1}{2}\left(D^{abc}+iC^{abc}\right)$. Taking the commutative limit $\theta\to 0$ in  both equation (\ref{M2-2-7b}) and (\ref{M2-2-8}) and using the antisymmetry of Grassmann variables $\overline{\psi}$ and $\psi$ leads to the usual definition of the current:
\begin{eqnarray*}
J_{\mu}^{a}=i{}\ \overline{\psi}_{\xi}\ T^{a}\ \psi_{\alpha}\left(\gamma_{\mu}P_{+}\right)^{\xi\alpha} =-{}\ \overline{\psi}_{\xi}^{c}C^{abc}\ \psi_{\alpha}^{b}\left(\gamma_{\mu}P_{+}\right)^{\xi\alpha},
\end{eqnarray*}
with $T^{a}$, $T^{a}_{\ bc}=iC^{abc}$, as the generators of U(N) in their adjoint representation.
\par
To study local gauge invariance and to determine the local chiral anomaly for NC-U(N), we consider the three-point function
\begin{eqnarray}\label{M2-2-9}
\Gamma_{\mu\lambda\nu}^{cfm}\left(x,y,z\right)=\bigg<\mbox{T}\ \left(J_{\mu}^{c}\left(x\right)J_{\lambda}^{f}\left(y\right)J_{\nu}^{m}\left(z\right)\right)\bigg>,
\end{eqnarray}
where $c,f,m=0,1,\cdots,N^{2}-1$ are U(N) indices. The contribution of diagram A and B, involving both planar and nonplanar phases, are given by
\begin{eqnarray}\label{M2-2-10}
\lefteqn{\bigg[\Gamma_{\mu\lambda\nu}^{cfm}\left(x,y,z\right)\bigg]_{A+B}=}\nonumber\\             
&&=\int\limits_{-\infty}^{+\infty}
\frac{d^{4}k_{2}}{\left(2\pi\right)^{4}}
\frac{d^{4}k_{3}}{\left(2\pi\right)^{4}}
\frac{d^{4}\ell}{\left(2\pi\right)^{4}}
e^{-i\left(k_{2}+k_{3}\right)x}\ e^{ik_{2}y}e^{ik_{3}z}
\bigg\{\mbox{Tr}\left(
D^{-1}\left(\ell-k_{3}\right)\gamma_{\mu}P_{+}D^{-1}\left(\ell+k_{2}\right)
\gamma_{\lambda}D^{-1}\left(\ell\right)\gamma_{\nu}\right)\nonumber\\
&&\times \sum\limits_{j=1}^{8}\big[{\cal{G}}_{A}^{cfm}\big]_{j}f_{A}^{j}\left(\ell\right)+[\left(k_{2},\lambda\right)\leftrightarrow\left(k_{3},\nu\right)]\sum\limits_{j=1}^{8}\big[{\cal{G}}_{B}^{cfm}\big]^{j}f_{B}^{j}\left(\ell\right)\bigg\},
\end{eqnarray}    
where the group factors $[{\cal{G}}_{A/B}^{cfm}]_{j}, j=1,\cdots,8$ and phases $f_{A/B}^{j}\left(\ell\right), j=1,\cdots,8$ are given in Table 1 and 2. 
\par 
Carrying out standard calculations it can be shown that quantum corrections do not affect the gauge invariance of noncommutative U(N) with adjoint matter fields. 
Next we calculate the anomaly of the theory. Going through standard current algebra analysis one can show that the anomalous part of $\partial^{\mu}\Gamma_{\mu\lambda\nu}$ is given by ${\cal{A}}_{\lambda\nu}^{cfm}\left(k_{2},k_{3},\theta\right)$, whose planar and nonplanar parts are treated separately. 
\par
Using the Feynman parametrization procedure and the properties of dimensionally regularized $\gamma_{5}$, the planar part of the anomaly reads: 
\begin{eqnarray}\label{M2-2-11}
\lefteqn{\hspace{-1.5cm}
\big[ {\cal{A}}_{\lambda\nu}^{cfm}\left(k_{2},k_{3};\theta\right)\big]_{\mbox{\small{pl.}}} =\frac{-i{}}{8\pi^{2}}\varepsilon_{\lambda\nu\alpha\beta}k_{2\alpha}k_{3\beta}
}\nonumber\\
&&\times\bigg\{e^{i\ k_{2}\times k_{3}}\left(\big[{\cal{G}}_{A}^{cfm}\big]_{1}+\big[{\cal{G}}_{B}^{cfm}\big]_{1}\right)+e^{-i\ k_{2}\times k_{3}}\left(\big[{\cal{G}}_{A}^{cfm}\big]_{8}+\big[{\cal{G}}_{B}^{cfm}\big]_{8}\right)\bigg\}.
\end{eqnarray}
After some group algebra, it can be shown that the group factors of both diagrams A and B cancel, i.e. 
$\big[{\cal{G}}_{A}^{cfm}\big]_{1}+\big[{\cal{G}}_{B}^{cfm}\big]_{1}=0$, and $\big[{\cal{G}}_{A}^{cfm}\big]_{8}+\big[{\cal{G}}_{B}^{cfm}\big]_{8}=0$.
Hence the planar part of the chiral anomaly vanishes.
\par 
For the nonplanar part of the anomaly, although the Feynman integrals including the nonplanar phases are naively convergent for finite values of the noncommutativity parameter $\theta$, they have to be  dimensionally regularized, as they are IR divergent for small $\theta$. Using 
\begin{eqnarray}\label{M2-2-12}
\{\left(\gamma_{\perp}\right)_{\xi}, \gamma_{\mu}\}=2\delta_{\xi\mu}\hspace{0.8cm}\mbox{for}\hspace{0.4cm}\mu=0,1,2,3, \hspace{1cm}\mbox{and}\ \ \ \big[\left(\gamma_{\perp}\right)_{\xi},\gamma_{5}\big]=0, 
\end{eqnarray}  
which lead to $
\mbox{Tr}\left(\gamma_{5}\left(\gamma_{\perp}\right)_{\xi}\gamma_{\lambda}\gamma_{\nu}\gamma_{\alpha}\right)=0$, and after some Dirac algebra we arrive at: 
\begin{eqnarray}\label{M2-2-14}
\big[{\cal{A}}_{\lambda\nu}^{cfm}\left(k_{2},k_{3};\theta\right)\big]_{\mbox{\small{n.-pl.}}}\hspace{-0.2cm}&=&-i{}\int\limits_{-\infty}^{+\infty}\frac{d^{D}\ell}{\left(2\pi\right)^{D}}\   
\Bigg\{\frac{
\mbox{Tr}\bigg[\left(\LS-\KS_{3}\right)\gamma_{5}\LS_{\perp}\left(\LS+\KS_{2}\right)\gamma_{\lambda}\LS\gamma_{\nu}\bigg]
}
{
\big[\left(\ell+k_{2}\right)^{2}\big]\big[\left(\ell-k_{3}\right)^{2}\big]\big[\ell^{2}\big] 
}\sum\limits_{j=2}^{7}\big[{\cal{G}}_{A}^{cfm}\big]_{j}f_{A}^{j}\left(\ell\right)\nonumber\\
&&+\big[\left(k_{2},\lambda\right)\leftrightarrow\left(k_{3},\nu\right)\big]
\sum\limits_{j=2}^{7}\big[{\cal{G}}_{B}^{cfm}\big]_{j}f_{B}^{j}\left(\ell\right)\Bigg\}.
\end{eqnarray}
Next, we use the Feynman parametrization and make the shift $\ell\to \ell+P_{a}$ with $P_{a}\equiv -k_{2}\alpha_{1}+k_{3}\alpha_{2}$ in the contribution of diagram $A$ and $\ell\to \ell+P_{b}$ with $P_{b}\equiv -P_{a}$ in the contribution of diagram $B$. Using the identity
$f_{A}^{j}\left(\ell\to-\ell+P_{a}\right)=f_{B}^{j}\left(\ell+P_{b}\right)$, we arrive at expressions of the form
\begin{eqnarray*}
I_{\eta\xi}&\equiv&\int\limits_{-\infty}^{+\infty}\frac{d^{D}\ell}{\left(2\pi\right)^{D}}\ \frac{\left(\ell_{||}\right)_{\eta}\left(\ell_{\perp}\right)_{\xi}e^{-2i\ell\times q}}{\left(\ell^{2}-\Delta\right)^{3}},\nonumber\\
\end{eqnarray*}
and
\begin{eqnarray}\label{M2-2-15}
I_{\xi}&\equiv&\int\limits_{-\infty}^{+\infty}\frac{d^{D}\ell}{\left(2\pi\right)^{D}}\ \frac{\left(\ell_{\perp}\right)_{\xi}e^{-2i\ell\times q}}{\left(\ell^{2}-\Delta\right)^{3}}.
\end{eqnarray}
As we will show in Appendix A, these integrals vanish for any dimension. The main contribution comes therefore from  
\begin{eqnarray}\label{M2-2-16}
\lefteqn{\hspace{-1cm}
\bigg[{\cal{A}}_{\lambda\nu}^{cfm}\left(k_{2},k_{3};\theta\right)\bigg]_{\mbox{\small{n.-pl.}}}= 8            {}\varepsilon_{\lambda\nu\alpha\beta}k_{2\alpha}k_{3\beta} 
}\nonumber\\
&&\hspace{-0.5cm}
\times
\sum\limits_{j=2}^{7} \left(\big[{\cal{G}}_{A}^{cfm}\big]_{j}+\big[{\cal{G}}_{B}^{cfm}\big]_{j} \right)\int\limits_{0}^{1}d\alpha_{1}\int\limits_{0}^{1-\alpha_{1}}d\alpha_{2}\int\limits_{-\infty}^{+\infty}\frac{d^{D}\ell}{\left(2\pi\right)^{D}}\ \frac{\ell_{\perp}^{2}f_{B}^{j}\left( \ell+k_{2}\alpha_{1}-k_{3}\alpha_{2}\right)
}
{\left(\ell^{2}-\Delta\right)^{3}},
\end{eqnarray}
where $\Delta\equiv -k_{2}^{2}\alpha_{1}\left(1-\alpha_{1}\right)-k_{3}^{2}\alpha_{2}\left(1-\alpha_{2}\right)-2k_{2}k_{3}\alpha_{1}\alpha_{2}$.
But, as it turns out, since the fermions are taken in the adjoint representation of the U(N) gauge group, the corresponding group factors in diagrams A and B cancel, i.e.
$\big[{\cal{G}}_{A}^{cfm}\big]_{j}+\big[{\cal{G}}_{B}^{cfm}\big]_{j}=0, \ \ \forall j=2,\cdots,7$. 
\par
Combining this with the previous result, we have shown that planar as well as nonplanar part of ${\cal{A}}_{\lambda\nu}^{cfm}$ vanish. Hence U(N) chiral gauge theory with adjoint matter fields turns out to be free of chiral gauge anomaly. 
\par
In the next section we will discuss the global symmetry and calculate the global anomaly of U(1) with a current which express the invariance of the corresponding action with respect to usual global symmetry transformation of the theory. 
\par\vskip0.2cm
\section{Planar and Nonplanar Contributions to Anomaly in NC-U(1)}
\setcounter{section}{3}
\setcounter{equation}{0}
\par\vskip0.2cm\noindent
{\it i) Classical Symmetries}
\par\vskip0.2cm\noindent
To check the global symmetry and classical current conservation laws for noncommutative U(1), let us briefly review the derivation of global currents in commutative gauge theories. 
\par
Consider the action $S\equiv \int d^{4}x \ {\cal{L}}\left(\psi^{\ell},\partial_{\mu}\psi^{\ell}\right)$ and a local transformation of fields 
\begin{eqnarray}\label{M3-1-1}
\delta_{\alpha}\psi^{\ell}=i\alpha\left(x\right) {\cal{F}}^{\ell}\left(x\right). 
\end{eqnarray}
If the action is not invariant under this local transformation, but is invariant under the corresponding global transformations, then its variation will have to be of the form
\begin{eqnarray}\label{M3-1-2}
\delta S=-\int d^{4}x J^{\mu}\left(x\right)\partial_{\mu}\alpha\left(x\right). 
\end{eqnarray}
Integrating by part leads to the conservation law 
\begin{eqnarray}\label{M3-1-3}
\partial^{\mu}J_{\mu}\left(x\right)=0,
\end{eqnarray}
giving $\frac{d}{dt}Q=0$ with $Q\equiv \int d^{3}x \ J_{0}\left(x\right)$. In commutative field theory, there is one such conserved current and one constant of motion for each independent infinitesimal symmetry transformation. This is the content of the Noether's first theorem. The current $J_{\mu}$ can be given explicitly, if the Lagrangian density is also invariant under the global symmetry transformation corresponding to (\ref{M3-1-1}). Then
\begin{eqnarray}\label{M3-1-4}
J_{\mu}\left(x\right)\equiv -i\frac{\partial {\cal{L}}}{\partial\left(\partial_{\mu}\psi^{\ell}\left(x\right)\right)}\ {\cal{F}}^{\ell}\left(x\right).
\end{eqnarray}
Back to the noncommutative gauge theory, let us first consider the action 
\begin{eqnarray}\label{M3-1-5}
S_{F}[\overline{\psi},\psi, A_{\mu}]=\int d^{D}x \bigg\{i\overline{\psi}_{\alpha}\left(x\right)\star\partial_{\mu}\psi_{\beta}\left(x\right)-g\ \overline{\psi}_{\alpha}\left(x\right)\star A_{\mu}\left(x\right)\star\psi_{\beta}\left(x\right)\bigg\}\left(\gamma_{\mu}\right)^{\alpha\beta},
\end{eqnarray}
for massless matter fields in the {\sl fundamental} representation. The corresponding Lagrangian density is invariant under global transformation
\begin{eqnarray}\label{M3-1-6}
\delta_{\alpha}\psi=i\alpha\psi\left(x\right), \hspace{1cm}\mbox{and}\hspace{1cm}\delta_{\alpha}\overline{\psi}=i\alpha\overline{\psi}\left(x\right).
\end{eqnarray}
Following the procedure presented above, it is now possible to define three different change of variables similar to (\ref{M3-1-1}) in noncommutative field theory
\begin{eqnarray}\label{M3-1-7}
\hspace{-1cm}I)\ \ \ \ \ \ \ \delta_{\alpha}\overline{\psi}_{\beta}=i\overline{\psi}_{\xi}\left(x\right)\star\alpha\left(x\right),\hspace{0.5cm}&\mbox{and}&\hspace{0.5cm}
\delta_{\alpha}\psi_{\xi}=i\alpha\left(x\right)\star\psi_{\beta}\left(x\right),\nonumber\\
\hspace{-1cm}II)\ \ \ \ \ \ \ \delta_{\alpha}\overline{\psi}_{\beta}=i\alpha\left(x\right)\star\overline{\psi}_{\xi}\left(x\right),\hspace{0.5cm}&\mbox{and}&\hspace{0.5cm}
\delta_{\alpha}\psi_{\xi}=i\psi_{\beta}\left(x\right)\star\alpha\left(x\right),\nonumber\\
\hspace{-1cm}III)\ \ \ \delta_{\alpha}\overline{\psi}_{\beta}=\frac{i}{2}\big\{\alpha\left(x\right),\overline{\psi}_{\xi}\left(x\right)\big\}_{\star},\hspace{0.5cm}&\mbox{and}&\hspace{0.5cm}
\delta_{\alpha}\psi_{\xi}=\frac{i}{2}\big\{\alpha\left(x\right),\psi_{\beta}\left(x\right)\big\}_{\star},
\end{eqnarray}
leading to three different vector currents
\newpage
\begin{eqnarray}\label{M3-1-8}
\hspace{-1cm}I)\ \ \ \ \ \ \ {\cal{J}}_{\mu}\left(x\right)&=&-i\psi_{\beta}\left(x\right)\star\overline{\psi}_{\alpha}\left(x\right)\left(\gamma_{\mu}\right)^{\alpha\beta},\nonumber\\
\hspace{-1cm}II)\ \ \ \ \ \ \ {\cal{J'}}_{\mu}\left(x\right)&=&+i\overline{\psi}_{\alpha}\left(x\right)\star\psi_{\beta}\left(x\right)\left(\gamma_{\mu}\right)^{\alpha\beta},\nonumber\\   
\hspace{-1cm}III)\ \ \ \ \ \ \ 
{\cal{J''}}_{\mu}\left(x\right)&=&-\frac{i}{2} \big[\psi_{\beta}\left(x\right),\overline{\psi}_{\alpha}\left(x\right)\big]_{\star}\left(\gamma_{\mu}\right)^{\alpha\beta}, 
\end{eqnarray}
respectively. Using the equation of motion 
\begin{eqnarray}\label{M3-1-8a}
\partial_{\mu}\overline{\psi}\gamma^{\mu}=ig\overline{\psi}\gamma^{\mu}\star A_{\mu},\hspace{1cm}\mbox{and}\hspace{1cm} 
\gamma^{\mu}\partial_{\mu}\psi=-igA_{\mu}\star\gamma^{\mu}\psi.
\end{eqnarray}
which follows from the corresponding Lagrangian to the action (\ref{M3-1-5}), it is possible to derive the following naive Ward identities
for both ${\cal{J}}_{\mu}(x)$ and ${\cal{J'}}_{\mu}(x)$
\begin{eqnarray}\label{M3-1-9}
\hspace{-0.8cm}D^{\mu}{\cal{J}}_{\mu}\left(x\right)=0,\hspace{0.5cm}&\mbox{and}&\hspace{0.5cm}
\partial^{\mu}{\cal{J'}}_{\mu}\left(x\right)=0,
\end{eqnarray}
where $D^{\mu}=\partial^{\mu}+ig\big[A^{\mu}\left(x\right),\cdot\big]_{\star}$ is the covariant derivative. The corresponding equation for  ${\cal{J''}}_{\mu}$ can be obtained by combining the above  equations and reads:  
$\partial^{\mu}{\cal{J''}}_{\mu}\left(x\right)+\frac{ig}{2}\big[A^{\mu}\left(x\right),{\cal{J}}_{\mu}\left(x\right)\big]_{\star}=0$.  
\par
To determine the local and global charachters of these currents and to define their conserved global charges, it is necessary to distinguish the different cases of spacelike and  timelike as well as lightlike noncommutativity tensor $\theta_{\mu\nu}$: 
\par\noindent
In the case of spacelike $\theta_{\mu\nu}$, making use of the definition of $\star$-product and assuming trivial boundary conditions for $f(x)$ and $g(x)$ we obtain
\begin{eqnarray}\label{M3-1-11}
\int d^{3}x\ f\left(x\right)\star g\left(x\right)=\int d^{3}x\  g\left(x\right)\star f\left(x\right).
\end{eqnarray}
Using then the relations given in Eq. (\ref{M3-1-9}), it can be shown that, all the currents from Eq. (\ref{M3-1-8}) lead to the same conserved charge $Q$, 
\begin{eqnarray}\label{M3-1-12}
Q\equiv \int d^{3}x\  J_{0}\left(x\right),\hspace{1cm}\forall J={\cal{J}}_{\mu}, {\cal{J'}}_{\mu}, {\cal{J''}}_{\mu}.   
\end{eqnarray}
If, however, only one of the components of $\theta_{\mu 0}, \mu=1,2,3$ is nonzero, the case for timelike or lightlike $\theta_{\mu\nu}$-tensor,  then Eq. (\ref{M3-1-2}) does not hold anymore. The relation $\partial^{\mu}{\cal{J'}}_{\mu}=0$, Eq. (\ref{M3-1-9}), leads, as before, to a conserved charge 
\begin{eqnarray}\label{M3-1-9a}
Q'=\int {\cal{J'}}_{0}(x) d^{3}x, \hspace{1cm}\mbox{with}\hspace{1cm}\partial^{0}Q'=0,
\end{eqnarray} 
which is different from $Q$ of Eq. (\ref{M3-1-12}). The vector current ${\cal{J}}_{\mu}(x)$ is therefore a {\sl local} one corresponding to a {\sl local} gauge transformation of the theory described by the {\sl fundamental} action of Eq. (\ref{M3-1-5}). It is the current which couples to $A_{\mu}$ in the action.  By assuming the usual gauge transformation for the gauge fields (\ref{M2-1-7}) this action will then be gauge invariant under {\sl local} transformation $(I)$ of Eq. (\ref{M3-1-7}). 
\par
To complete the list of properties satisfied by different currents of noncommutative field theory, we add that the same continuity equation 
\begin{eqnarray}\label{M3-1-13}
\int\ \partial^{\mu}J_{\mu}\left(x\right)\  dx_{\theta}=0\hspace{1cm}\forall J={\cal{J}}_{\mu}, {\cal{J'}}_{\mu}, {\cal{J''}}_{\mu},   
\end{eqnarray}   
is valid for any arbitrary noncommutativity tensor $\theta_{\mu\nu}$. This can be shown by integrating equations (\ref{M3-1-9}) over all noncommutative space-time coordinates, i.e. by definition over $dx_{\theta}$, and using the definition of $\star$-product.
\par
Similarly there are three different axial vector currents 
\begin{eqnarray}\label{M3-1-8b}
\hspace{-1cm}I)\ \ \ \ \ \ \ {\cal{J}}_{\mu,5}\left(x\right)&=&-i\psi_{\beta}\left(x\right)\star\overline{\psi}_{\alpha}\left(x\right)\left(\gamma_{\mu}\gamma_{5}\right)^{\alpha\beta},\nonumber\\
\hspace{-1cm}II)\ \ \ \ \ \ \ {\cal{J'}}_{\mu,5}\left(x\right)&=&+i\overline{\psi}_{\alpha}\left(x\right)\star\psi_{\beta}\left(x\right)\left(\gamma_{\mu}\gamma_{5}\right)^{\alpha\beta},\nonumber\\   
\hspace{-1cm}III)\ \ \ \ \ \ \ 
{\cal{J''}}_{\mu,5}\left(x\right)&=&-\frac{i}{2} \big[\psi_{\beta}\left(x\right),\overline{\psi}_{\alpha}\left(x\right)\big]_{\star}\left(\gamma_{\mu}\gamma_{5}\right)^{\alpha\beta}, 
\end{eqnarray}
resulting from the global axial invariance of the action (\ref{M3-1-5}) with respect to global axial transformation $\partial_{\alpha}\psi=i\alpha\gamma_{5}\psi$. They are associated with local transformation similar to those given in Eq. (\ref{M3-1-7}). Similar conservation equations hold also for these axial vector currents.  
\par
The above facts about vector currents are not only valid for the action (\ref{M3-1-5}), where the matter fields in the fundamental representation are considered, they are also true when the matter fields are in the anti-fundamental or adjoint representation. 
\par 
In Ref.\cite{neda3}, the global anomaly corresponding to the axial vector current ${\cal{J'}}_{\mu,5}$ was calculated in a theory formulated with anti-fundamental matter fields. It was shown that, only planar diagrams contribute to axial anomaly, that could be expressed by the usual commutative anomaly with the ordinary products replaced with noncommutative $\star$-products. 
\par 
We will now calculate the anomaly corresponding to the axial vector current ${\cal{J}}_{\mu,5}$ and ${\cal{J'}}_{\mu,5}$, in a theory with fundamental matter fields, and find nonplanar contributions.  
\par\vskip0.5cm\noindent
{\it ii) Anomaly and Nonplanar Diagrams in NC-U(1)}
\par\vskip0.2cm\noindent
Let us consider the  theory described by the action (\ref{M3-1-5}) and its corresponding current ${\cal{J}}_{\mu,5}(x)$. To find the quantum corrections to the classical equation satisfied by this current, $D^{\mu}{\cal{J}}_{\mu,5}(x)=0$, the three-point function    
\begin{eqnarray}\label{M3-2-1}
\Gamma_{\mu\lambda\nu}\left(x,y,z\right)\equiv\bigg<T\left(
{\cal{J}}_{\mu,5}\left(x\right){\cal{J}}_{\lambda}\left(y\right){\cal{J}}_{\nu}\left(z\right)\right)
\bigg>,
\end{eqnarray}
has to be considered. The Feynman integrals corresponding to the diagrams of Fig. \ref{Fig1} are given again by (\ref{M2-1-10}), where now $P_{+}$ has to be replaced by $\gamma_{5}$. Note that here there appear only {\sl planar} phase factors   
\begin{eqnarray}
F_{a}=e^{i\ k_{2}\times k_{3}}, \hspace{1cm} \mbox{and}\hspace{1cm}
F_{b}=e^{-i\ k_{2}\times k_{3}}.
\end{eqnarray}
After a calculation similar to what which was performed in Ref. \cite{neda3}, the axial anomaly reads
\begin{eqnarray}\label{planar}
\bigg<D^{\mu}{\cal{J}}_{\mu,5}(x)\bigg>=-\frac{g^{2}}{16\pi^{2}}\varepsilon^{\lambda\nu\alpha\beta}F_{\lambda\nu}(x)\star F_{\alpha\beta}(x).
\end{eqnarray} 
To obtain this result it is necessary to consider also the contributions of the diagrams shown in Fig. \ref{Fig2}. Both diagrams  turn out to be planar, so that no new effects will occur in this case. To obtain a $\star$-gauge invariant result an integration over noncommutative space-time coordinates has to be performed. This solution,  however, seems to pick up only the zero momentum mode in the Fourier transformed of the above result. To obtain all Fourier modes, a straight Wilson line, with a length proportional to the noncommutativity parameter $\theta$,  must be attached to $F_{\mu\nu}$ operators\cite{gross}. In the expansion of the Wilson line in terms of the external gauge field $A_{\mu}$, new terms appear which are higher order than those of Figs. \ref{Fig1} and \ref{Fig2}. This is in contradistinction with the Adler-Bardeen's theorem\cite{adler} in commutative non-Abelian gauge theory, which states that no more diagrams than those in Figs. \ref{Fig1} and \ref{Fig2} are needed to obtain a gauge invariant result for the anomaly.
\par
As it was argued before, the noncommutative gauge theory has another conserved axial currents ${\cal{J'}}_{\mu,5}$. To calculate the quantum corrections consider
\begin{eqnarray}
\tilde{\Gamma}_{\mu\lambda\nu}\left(x,y,z\right)\equiv\bigg<T\left(
{\cal{J'}}_{\mu,5}\left(x\right){\cal{J}}_{\lambda}\left(y\right){\cal{J}}_{\nu}\left(z\right)\right)
\bigg>.
\end{eqnarray}
Again the Feynman integrals corresponding to the diagrams of Fig. \ref{Fig1} are given by (\ref{M2-1-10}), where now $P_{+}$ is replaced by $\gamma_{5}$. Note that here there appear only {\sl nonplanar} phase factors depending on the integration loop momenta $\ell$  
\begin{eqnarray}\label{phases}
F_{a}(\ell)=e^{i\ k_{2}\times k_{3}+2i\ell\times (k_{2}+k_{3})}, \hspace{1cm} \mbox{and}\hspace{1cm}
F_{b}(\ell)=e^{-i\ k_{2}\times k_{3}+2i\ell\times (k_{2}+k_{3})}.
\end{eqnarray}
In order to study a possible UV/IR mixing in the nonplanar diagrams we have to go through an appropriate regularization scheme. We do both dimensional and Pauli-Villars regularizations. 
\par
Using again the 't Hooft procedure, with
\begin{eqnarray*}
(\KS_{2}+\KS_{3})\gamma_{5}=-\gamma_{5}D(\ell+k_{3})-D(\ell-k_{2})\gamma_{5}+2\gamma_{5}\LS_{\perp},
\end{eqnarray*} 
in the contribution of diagram (a) Fig. \ref{Fig1} to $\partial^{\mu}\tilde{\Gamma}_{\mu\lambda\nu}(x,y,z)$ and 
\begin{eqnarray*}
(\KS_{2}+\KS_{3})\gamma_{5}=-\gamma_{5}D(\ell+k_{2})-D(\ell-k_{3})\gamma_{5}+2\gamma_{5}\LS_{\perp},
\end{eqnarray*}
in the contribution of diagram (b) Fig. \ref{Fig1}, to obtain 
\begin{nedalph}\label{M3-2-3a}
\partial^{\mu}\tilde{\Gamma}_{\mu\lambda\nu}=
\int\limits_{-\infty}^{+\infty}
\frac{d^{4}k_{2}}{\left(2\pi\right)^{4}}
\frac{d^{4}k_{3}}{\left(2\pi\right)^{4}}\ e^{-i\left(k_{2}+k_{3}\right) x}
\ e^{ik_{2}y}\ e^{ik_{3}z}
 \bigg[{\cal{A}}_{\lambda\nu}\left(k_{2},k_{3};\theta\right)+{\cal{R}}_{\lambda\nu}\left(k_{2},k_{3};\theta\right)\bigg],
\end{eqnarray}
where 
\begin{eqnarray}\label{M3-2-3b}
{\cal{A}}_{\lambda\nu}\left(k_{2},k_{3};\theta\right)
&\equiv & -2i\int\limits_{-\infty}^{+\infty}\frac{d^{D}\ell}{\left(2\pi\right)^{D}}
\Bigg[\mbox{Tr}\left( D^{-1}\left(\ell-k_{3}\right)\gamma_{5}\LS_{\perp}D^{-1}\left(\ell+k_{2}\right)\gamma_{\lambda}D^{-1}\left(\ell\right)\gamma_{\nu}\right)\ F_{a}\left(\ell\right)\Bigg]\nonumber\\
&&+\big[\left(k_{2},\lambda\right)\leftrightarrow \left(k_{3},\nu\right)\big]F_{b}\left(\ell\right),
\end{eqnarray}
and  
\begin{eqnarray}\label{M3-2-3c}
\lefteqn{{\cal{R}}_{\lambda\nu}\left(k_{2},k_{3};\theta\right)\equiv}\nonumber\\
&&\hspace{-0.6cm}\equiv +i\int\limits_{-\infty}^{+\infty}\frac{d^{D}\ell}{\left(2\pi\right)^{D}}
\Bigg\{\bigg[
\mbox{Tr}\left(D^{-1}\left(\ell-k_{3}\right)\gamma_{5}\gamma_{\lambda}D^{-1}\left(\ell\right)\gamma_{\nu}\right)+
\mbox{Tr}\left(\gamma_{5}D^{-1}\left(\ell+k_{2}\right)\gamma_{\lambda}D^{-1}\left(\ell\right)\gamma_{\nu}\right)\bigg]
F_{a}\left(\ell\right)\nonumber\\
&&\hspace{-0.5cm}+\bigg[
\mbox{Tr}\left(D^{-1}\left(\ell-k_{2}\right)\gamma_{5}\gamma_{\nu}D^{-1}\left(\ell\right)\gamma_{\lambda}\right)+
\mbox{Tr}\left(\gamma_{5}D^{-1}\left(\ell+k_{3}\right)\gamma_{\nu}D^{-1}\left(\ell\right)\gamma_{\lambda}\right)
\bigg]F_{b}\left(\ell\right)\Bigg\}.
\end{nedalph}
After appropriate shifts of integration variable the rest terms  ${\cal{R}}_{\lambda\nu}$ can be shown to vanish. The anomalous part of $\partial^{\mu}\tilde{\Gamma}_{\mu\lambda\nu}$ is therefore given by ${\cal{A}}_{\lambda\nu}$, which consits only of nonplanar part. After a straightforward calculation this is given by
\begin{eqnarray}\label{M3-2-9}
{\cal{A}}_{\lambda\nu}= -16\varepsilon_{\lambda\nu\alpha\beta}k_{2}^{\alpha}k_{3}^{\beta}
\int\limits_{0}^{1}d\alpha_{1}\int\limits_{0}^{1-\alpha_{1}}d\alpha_{2}\int\limits_{-\infty}^{+\infty}\frac{d^{D}\ell}{\left(2\pi\right)^{D}}\ \frac{\ell_{\perp}^{2}\ F_{b}\left(\ell+k_{2}\alpha_{1}-k_{3}\alpha_{2}\right)
}
{\left(\ell^{2}+\Delta\right)^{3}},
\end{eqnarray}
where $\Delta\equiv k_{2}^{2}\alpha_{1}\left(1-\alpha_{1}\right)+k_{3}^{2}\alpha_{2}\left(1-\alpha_{2}\right)+2k_{2}k_{3}\alpha_{1}\alpha_{2}$. 
To perform the $\ell$-integration we use Eqs. (\ref{B8}) and (\ref{B18}) [Appendix A],  
\begin{nedalph}\label{M3-2-10a}
I_{1}&\equiv&\int\limits_{-\infty}^{+\infty}\frac{d^{D}\ell}{\left(2\pi\right)^{D}}\ \frac{\ell_{\perp}^{2}e^{\pm 2i\ell\times q}}{\left(\ell^{2}+\Delta\right)^{3}}=\frac{\left(D-4\right)}{D}
\int\limits_{-\infty}^{+\infty}\frac{d^{D}\ell}{\left(2\pi\right)^{D}}\ \frac{\ell^{2}e^{\pm 2i\ell\times q}}{\left(\ell^{2}+\Delta\right)^{3}}\nonumber\\
&=&\frac{1}{\ln\Lambda^{2}}\bigg[\frac{1}{16\pi^{2}}\ {\cal{E}}_{1}\left(q,\Delta\right)-\frac{q\circ q}{128\pi^{2}}\ {\cal{E}}_{2}\left(q,\Delta\right)\bigg],
\end{eqnarray}
where   
\begin{eqnarray}\label{M3-2-10b}
{\cal{E}}_{n}\left(q,\Delta\right)\equiv\int\limits_{0}^{\infty}\frac{d\rho}{\rho^{n}}\ \mbox{exp}\bigg[-\rho\ \Delta-\frac{q\circ q}{4}\ \frac{1}{\rho}\bigg].  
\end{nedalph}
with $q\circ q\equiv -q_{\sigma}\theta_{\sigma\eta}\theta_{\eta\xi}\ q_{\xi}$. In order to  compare with the cut-off regularization method we have replaced the factor $D-4$ before the integral on the first line of the Eq. (\ref{M3-2-10a}) by $\frac{1}{\ln\Lambda^{2}}$. Note that in Eq. (\ref{M3-2-10a}) the integrals with the phases $e^{+2i\ell\times q}$ or $e^{-2i\ell\times q}$ yield the same result, because $I_{1}$ is even under  $\theta\to -\theta$.  
\par
The functions ${\cal{E}}_{n}\left(q,\Delta\right), n=1,2$, for large $\Lambda_{eff}$ can be approximated by\cite{minwalla,hayakawa2}
\begin{eqnarray}\label{M3-2-11}
{\cal{E}}_{1}\left(q,\Delta\right)&=&\int\limits_{0}^{\infty}\frac{d\rho}{\rho}\mbox{exp}\bigg[-\rho\ \Delta-\frac{1}{\Lambda^{2}_ {\mbox{\small{eff.}}}} \ \frac{1}{\rho}\bigg]= +\ln\frac{\Lambda^{2}_ {\mbox{\small{eff.}}}}  {\Delta}+{\cal{O}}\left(1\right).\nonumber\\  
{\cal{E}}_{2}\left(q,\Delta\right)&=&\int\limits_{0}^{\infty}\frac{d\rho}{\rho^{2}}\mbox{exp}\bigg[-\rho\ \Delta-\frac{1}{\Lambda^{2}_ {\mbox{\small{eff.}}}}  \ \frac{1}{\rho}\bigg]= \Lambda^{2}_{\mbox{\small{eff.}}}-\Delta\ln\frac{\Lambda^{2}_ {\mbox{\small{eff.}}}} {\Delta}+{\cal{O}}\left(1\right),  
\end{eqnarray}
with
\begin{eqnarray}\label{M3-2-12}
\frac{1}{\Lambda^{2}_{\mbox{\small{eff.}}}}\equiv \frac{1}{\Lambda^{2}}+\frac{q\circ q}{4}.
\end{eqnarray}
Now putting the expression from Eq. (\ref{M3-2-10a}) in the Eq. (\ref{M3-2-9}) we arrive at:
\begin{eqnarray}\label{M3-2-13a}
{\cal{A}}_{\lambda\nu}\left(k_{2},k_{3};\theta,\Lambda\right)&=&-\frac{2}{\pi^{2}}\varepsilon_{\lambda\nu\alpha\beta}k_{2}^{\alpha}k_{3}^{\beta}\int\limits_{0}^{1}d\alpha_{1}\int\limits_{0}^{1-\alpha_{1}}d\alpha_{2} \cos\big[k_{2}\times k_{3}(1-2\alpha_{1}-2\alpha_{2})\big],\nonumber\\
&&\times \frac{1}{\ln\Lambda^{2}}
\bigg[{\cal{E}}_{1}\left(k_{1},\Delta\right)- \frac{k_{1}\circ k_{1}}{8}\  {\cal{E}}_{2}\left(k_{1},\Delta\right)\bigg],
\end{eqnarray}
where $k_{1}\equiv\left(k_{2}+k_{3}\right)$. Replacing then this result in Eq. (\ref{M3-2-3a}) and using the relation 
\begin{eqnarray*}
\big<{\cal{J'}}_{\mu,5}(x)\big>\equiv \frac{1}{2}\int d^{4}y\  d^{4}z\  \tilde{\Gamma}_{\mu\lambda\nu}(x,y,z)\  A^{\lambda}(y)\ A^{\nu}(z),
\end{eqnarray*}
we arrive at
\begin{eqnarray}\label{M3-2-14}
\lefteqn{\bigg<\partial^{\mu}{\cal{J'}}_{\mu,5}(x)\bigg>= -\frac{1}{\pi^{2}}\varepsilon_{\lambda\nu\alpha\beta}\
\int\limits_{-\infty}^{+\infty}
\frac{d^{4}k_{2}}{\left(2\pi\right)^{4}}\ 
k_{2}^{\alpha}\ A_{\lambda}(k_{2})\ e^{-ik_{2} x}\ 
\int\limits_{-\infty}^{+\infty}
\frac{d^{4}k_{3}}{\left(2\pi\right)^{4}}\ 
k_{3}^{\beta}\ A_{\nu}(k_{3})e^{-ik_{3}x}
}\nonumber\\
&&\times \int\limits_{0}^{1}d\alpha_{1}\int\limits_{0}^{1-\alpha_{1}}d\alpha_{2} \cos\big[k_{2}\times k_{3}(1-2\alpha_{1}-2\alpha_{2})\big]\nonumber\\
&&\times \frac{1}{\ln\Lambda^{2}}\Bigg[\left(
\ln\frac{1}{\frac{1}{\Lambda^{2}}+\frac{\left(k_{1}\circ k_{1}\right)}{4}}-\ln\Delta\right)-
\frac{\left(k_{1}\circ k_{1}\right)}{8}\left(\frac{1}{\frac{1}{\Lambda^{2}}+\frac{\left(k_{1}\circ k_{1}\right)}{4}}-\Delta\ln\frac{1}{\frac{1}{\Lambda^{2}}+\frac{\left(k_{1}\circ k_{1}\right)}{4}}+\Delta\ln\Delta\right)\Bigg],\nonumber\\
\end{eqnarray}
where the integration over $y$ and $z$ is performed and ${\cal{E}}_{i}(q,\Delta), i=1,2$ are replaced from Eq. (\ref{M3-2-11}). To discuss the UV/IR mixing effect, let us consider two different limits:
\begin{enumerate}
\item[{\it i)}] For $\frac{k_{1}\circ k_{1}}{4}>>\frac{1}{\Lambda^{2}}$, the anomaly vanishes due to the factor $\frac{1}{\ln\Lambda^{2}}$ in front of the expression on the third line of Eq. (\ref{M3-2-14}). Equivalently, we could first take the limit $\Lambda\to\infty$ (or $D\to 4$) and then $\theta\to 0$. 
\item[{\it ii)}] Considering the case $\frac{k_{1}\circ k_{1}}{4}<<\frac{1}{\Lambda^{2}}$ a finite anomaly arises due to IR singularity. Taking first the limit $\theta\to 0$ and then  $\Lambda\to \infty$ (or $D\to 4$) shows that there is only one nonvanishing contribution from the factor   
\begin{eqnarray}\label{M3-2-14b}
\lim\limits_{\Lambda\to\infty}\frac{1}{\ln\Lambda^{2}}\ 
\ln\frac{1}{\frac{1}{\Lambda^{2}}}=1,
\end{eqnarray}
on the third line of the Eq. (\ref{M3-2-14}). All other terms can be neglected due to the small $\theta$-parameter. After integrating over $\alpha_{1}$ and $\alpha_{2}$, we arrive at:
\end{enumerate}
\begin{eqnarray}\label{M3-2-18}
\lefteqn{\bigg<\partial^{\mu}{\cal{J'}}_{\mu,5}(x)\bigg>=}\nonumber\\
&=& -\frac{1}{2\pi^{2}}\varepsilon_{\lambda\nu\alpha\beta}\
\hspace{-.3cm}\int\limits_{\frac{k_{1}\circ k_{1}}{4}<<\frac{1}{\Lambda^{2}}}
\frac{d^{4}k_{2}}{\left(2\pi\right)^{4}}\ 
\frac{d^{4}k_{3}}{\left(2\pi\right)^{4}}\ 
\ \partial^{\lambda}A^{\nu}(k_{2})\ e^{-ik_{2} x}\ 
\frac{\sin\left(k_{2}\times k_{3}\right)}{k_{2}\times k_{3}}
\ e^{-ik_{3}x}\ 
\partial^{\alpha}A^{\beta}(k_{3}).
\end{eqnarray}
Defining now a new (generalized) $\star$-product 
\begin{eqnarray}\label{M3-2-19}
f\left(x\right)\star' g\left(x\right)&\equiv& f\left(x\right)\frac{\sin\left(\frac{\theta^{\mu\nu}}{2}\stackrel{\leftarrow}{\partial}_{\mu}\stackrel{\rightarrow}{\partial}_{\nu}\right)}{\frac{\theta^{\mu\nu}}{2}\stackrel{\leftarrow}{\partial}_{\mu}\stackrel{\rightarrow}{\partial}_{\nu}}g\left(x\right)\nonumber\\
&=&f\left(x\right)g\left(x\right)-\frac{1}{3!\times 4}\theta^{\mu\nu}\theta^{\rho\eta}\partial_{\mu}\partial_{\rho}f\left(x\right)\partial_{\nu}\partial_{\eta}g\left(x\right)+\cdots, 
\end{eqnarray}
equation (\ref{M3-2-18}) may be written in a more compact form as
\begin{eqnarray}\label{M3-2-20}
\bigg<\partial^{\mu}{\cal{J'}}_{\mu,5}(x)\bigg>
= -\frac{1}{2\pi^{2}}\varepsilon_{\lambda\nu\alpha\beta}\
\hspace{-.3cm}\int\limits_{\frac{k_{1}\circ k_{1}}{4}<<\frac{1}{\Lambda^{2}}}
\frac{d^{4}k_{2}}{\left(2\pi\right)^{4}}\ 
\frac{d^{4}k_{3}}{\left(2\pi\right)^{4}}\ 
\ \partial^{\lambda}A^{\nu}(k_{2})\ e^{-ik_{2} x}\star' 
\ e^{-ik_{3}x}\ 
\partial^{\alpha}A^{\beta}(k_{3}).
\end{eqnarray}
A product of this form appear also in the literature\cite{garousi,liustar,reststar}. Adding now the contribution of diagrams from Fig. \ref{Fig2} we obtain:
 \begin{eqnarray}\label{M3-2-21}
\bigg<\partial^{\mu}{\cal{J'}}_{\mu,5}(x)\bigg>
= -\frac{1}{8\pi^{2}}\varepsilon_{\lambda\nu\alpha\beta}\
\hspace{-.3cm}\int\limits_{\frac{k_{1}\circ k_{1}}{4}<<\frac{1}{\Lambda^{2}}}
\frac{d^{4}k_{2}}{\left(2\pi\right)^{4}}\ 
\frac{d^{4}k_{3}}{\left(2\pi\right)^{4}}\ 
\ F^{\lambda\nu}(k_{2})\ e^{-ik_{2} x}\star' 
\ e^{-ik_{3}x}\ 
F^{\alpha\beta}(k_{3}).
\end{eqnarray}
As we have seen, taking the limit $\theta\to 0$ and $\Lambda\to \infty$ do not commute. We conclude therefore that the finite anomaly from nonplanar diagrams is due to UV/IR mixing \cite{minwalla}.  
\par
In the  Pauli-Villars regularization method, introducing the regulator mass $M$ we obtain:
\begin{eqnarray}\label{M3-2-22}
\bigg[\partial^{\mu}\tilde{\Gamma}_{\mu\lambda\nu}\bigg]_{\mbox{\small{reg.}}}&=&\lim\limits_{M\to \infty}-32M^{2}\varepsilon_{\lambda\nu\alpha\beta}\int\limits_{-\infty}^{+\infty}
\frac{d^{4}k_{2}}{\left(2\pi\right)^{4}}
\frac{d^{4}k_{3}}{\left(2\pi\right)^{4}}\ e^{-i\left(k_{2}+k_{3}\right) x}
\ e^{ik_{2}y}\ e^{ik_{3}z}k_{2}^{\alpha}k_{3}^{\beta}\nonumber\\
&&\times\int\limits_{0}^{1}d\alpha_{1}\int\limits_{0}^{1-\alpha_{2}}d\alpha_{2}\cos\big[k_{2}\times k_{3}(1-2(\alpha_{1}+\alpha_{2}))\big] \int\limits_{-\infty}^{+\infty}\frac{d^{4}\ell}{(2\pi)^{4}}\ \frac{e^{2i\ell\times (k_{2}+k_{3})}}{(\ell^{2}-\Delta_{M})^{3}}.
\end{eqnarray}  
Here, $\Delta_{M}\equiv M^{2}-\alpha_{1}(1-\alpha_{1})k_{2}^{2}-\alpha_{2}(1-\alpha_{2})k_{3}^{2}-2\alpha_{1}\alpha_{2}k_{2}k_{3}$. 
The $\ell$ integration can be performed using the same manipulations introduced in Appendix A:
\begin{eqnarray}\label{M3-2-23}
\int\limits_{-\infty}^{+\infty} \frac{d^{4}\ell}{(2\pi)^{4}}\ \frac{e^{2i\ell\times k_{1}}}{(\ell^{2}-\Delta_{M})^{3}}=\frac{1}{16\pi^{2}}\int\limits_{0}^{\infty}d\rho\ e^{-\rho\Delta_{M}-\frac{k_{1}\circ k_{1}}{4\rho}}.
\end{eqnarray}
The Pauli-Villars regulated $\partial^{\mu}\tilde{\Gamma}_{\mu\lambda\nu}$ from Eq. (\ref{M3-2-22}) is  therefore given by:
\begin{eqnarray}\label{M3-2-24}
\lefteqn{\hspace{-1.5cm}\bigg[\partial^{\mu}\tilde{\Gamma}_{\mu\lambda\nu}(x,y,z)\bigg]_{\mbox{\small{reg.}}}=\lim\limits_{M\to \infty}-\frac{2}{\pi^{2}}M^{2}\varepsilon_{\lambda\nu\alpha\beta}\int\limits_{-\infty}^{+\infty}
\frac{d^{4}k_{2}}{\left(2\pi\right)^{4}}
\frac{d^{4}k_{3}}{\left(2\pi\right)^{4}}\ e^{-i\left(k_{2}+k_{3}\right) x}
\ e^{ik_{2}y}\ e^{ik_{3}z}k_{2}^{\alpha}k_{3}^{\beta}
}\nonumber\\ 
&&\times\int\limits_{0}^{1}d\alpha_{1}\int\limits_{0}^{1-\alpha_{2}}d\alpha_{2}\cos\big[k_{2}\times k_{3}(1-2(\alpha_{1}+\alpha_{2}))\big]\int\limits_{0}^{\infty}d\rho\ e^{-\rho M^{2}-\frac{k_{1}\circ k_{1}}{4\rho}},
\end{eqnarray}
where for large enough Pauli-Villars mass, $\Delta_{M}$ appearing on the r.h.s. of Eq. (\ref{M3-2-23}) is replaced by $M^{2}$.\footnote{The same expression appears also in \cite{ken} and leads to the same $\star'$ structure which we have obtained using dimensional regularization in the first version of this paper (hep-th/0009233).} 
Using now the approximation 
\begin{eqnarray*}
\int\limits_{0}^{\infty}d\rho\ e^{-\rho\ M^{2}-\frac{k_{1}\circ k_{1}}{4\rho}}\approx \frac{1}{M^{2}}+\frac{k_{1}\circ k_{1}}{4}(-1+2\gamma_{E}+\ln \frac{M^{2}(k_{1}\circ k_{1})}{4}),
\end{eqnarray*}
where $\gamma_{E}$ is the Euler number, in the above equation we arrive at:
\begin{eqnarray}\label{M3-2-25}
\lefteqn{\hspace{0cm}\bigg[\partial^{\mu}\tilde{\Gamma}_{\mu\lambda\nu}(x,y,z)\bigg]_{\mbox{\small{reg.}}}=\lim\limits_{M\to \infty}-\frac{2}{\pi^{2}}M^{2}\varepsilon_{\lambda\nu\alpha\beta}\int\limits_{-\infty}^{+\infty}
\frac{d^{4}k_{2}}{\left(2\pi\right)^{4}}
\frac{d^{4}k_{3}}{\left(2\pi\right)^{4}}\ e^{-i\left(k_{2}+k_{3}\right) x}
\ e^{ik_{2}y}\ e^{ik_{3}z}k_{2}^{\alpha}k_{3}^{\beta}
}\nonumber\\ 
&&\times\int\limits_{0}^{1}d\alpha_{1}\int\limits_{0}^{1-\alpha_{2}}d\alpha_{2}\cos\big[k_{2}\times k_{3}(1-2(\alpha_{1}+\alpha_{2}))\big] \left(\frac{1}{M^{2}}+\frac{k_{1}\circ k_{1}}{4}(-1+2\gamma_{E}+\ln \frac{M^{2}(k_{1}\circ k_{1})}{4})\right).\nonumber\\
\end{eqnarray}
It is easily seen, that the limits $M\to\infty$ and $\theta\to 0$ do not commute. If we take the limit $M\to\infty$ first and then $\theta\to 0$ the anomaly from nonplanar diagrams vanishes. If we consider the case $\frac{k_{1}\circ k_{1}}{4}<<\frac{1}{M^{2}}$, or equivalently take first the limit $\theta\to 0$ and then $M\to\infty$, factor $M^{2}$ cancels on the r.h.s. of Eq. (\ref{M3-2-25}) and a finite anomaly appears. Then after performing the integration over parameters $\alpha_{1}$ and $\alpha_{2}$, we arrive at the same result for the anomaly of nonplanar diagrams as calculated by dimensional regularization [Eq. (\ref{M3-2-18})]. Using the definition of generalized $\star$ product, Eq. (\ref{M3-2-19}), we obtain first Eq. (\ref{M3-2-20}) and finally after considering the contribution from the next higher loop orders (diagrams of Fig. \ref{Fig2}), we arrive at the result given in Eq. (\ref{M3-2-21}). 
\par
Note that due to the $\star'$-product, which is commutative but not associative, the last result from Eq. (\ref{M3-2-21}) is not $\star$-gauge invariant. To obtain a gauge invariant result, it is necessary to connect the operators $F_{\mu\nu}$ to open Wilson-line using an appropriate path ordering\cite{liustar}.   
\par
\section{Conclusion}
\noindent
In this paper we have studied the effect of nonplanar diagrams on the gauge and global anomalies of noncommutative gauge theories. In the first part we studied the chiral gauge anomaly of U(1) and U(N) theories, with adjoint matter fields and found out, that both theories are free of such anomalies. 
\par 
In the second part we studied global symmetries of the U(1) theory with fundamental matter field. Here we found a novel result, which can be extended to anti-fundamental and adjoint cases. Following the standard methods from commutative field theory, three different vector and axial vector currents were derived from three different local change of variables. We have shown that, for spacelike noncommutativity tensor $\theta_{\mu\nu}$, all currents expressed the same global symmetry of the action and led to the same conserved charge. For timelike or lightlike noncommutativity, however, there was only one global vector current ${\cal{J'}}_{\mu}$ satisfying the Abelian naive Ward identity $\partial^{\mu}{\cal{J'}}_{\mu}(x)=0$. The other vector current ${\cal{J}}_{\mu}$, which is at the same time the current coupled to the gauge field in the action, is associated with $D^{\mu}{\cal{J}}_{\mu}(x)=0$. Both axial vector currents ${\cal{J}}_{\mu,5}$ and ${\cal{J'}}_{\mu,5}$ were shown to be anomalous. Using the result which was obtained in Ref.\cite{neda3}, we have shown that in a theory containing fundamental matter field, the global anomaly of ${\cal{J}}_{\mu,5}$ arises from planar diagrams and has the same structure as in the commutative case with noncommutative $\star$-products replacing the usual products. 
\par\noindent
The corresponding anomaly to ${\cal{J'}}_{\mu,5}$, however, arises from nonplanar diagrams and has a structure involving new,  generalized $\star$-product. The finite anomaly from nonplanar diagrams is a consequence of the UV/IR mixing\cite{minwalla}.\footnote{Recently, our result is also confirmed by Intrilagator and Kumar \cite{ken} in the more general context of string theory.} Using the same methods, it can easily be shown, that the third current of the theory ${\cal{J''}}_{\mu,5}$ is also anomalous. Its anomaly arises from planar {\it and} nonplanar digrams and involves both $\star$- as well as $\star'$-products. 
\par
Note that the final results of the anomaly from planar and nonplanar diagrams are not $\star$-gauge invariant. To restore the $\star$-gauge invariance, it is necessary to attach both operators $F_{\mu\nu}$ and its dual to open, straight Wilson lines with a length proportional to the noncommutativity parameter $\theta$.\cite{gross,garousi,liustar,reststar}.  The zeroth order of the expansion of this Wilson line in orders of external gauge fields, corresponds to the results which we have reproduced by perturbative calculation of the diagrams from Figs. \ref{Fig1} and \ref{Fig2}. Only after considering an infinite number of higher loop diagrams, arising from all higher order terms in the expansion of the Wilson lines, a gauge invariant result for the anomaly (from planar as well as nonplanar diagrams) can be achieved. This is in contradistinction with Adler-Bardeen's theorem\cite{adler} in the commutative gauge theories, which states that no more diagrams than those of Figs. \ref{Fig1} and \ref{Fig2} are needed to obtain a gauge invariant result for the anomaly of non-Abelian gauge theories.   
\par\vskip0.5cm
\section{Acknowledgments}
\noindent
We would like to thank H. Arfaei and H. Yavartanoo for discussions.
\vskip0.5cm
\par\noindent{\bf Note added:} After this work (hep-th/0009233) several studies on many different aspects of generalized $\star$-product appeared in Refs.\cite{garousi,liustar,reststar}. 
 
\newpage
\begin{appendix}
\noindent
The Jacobi Identities between the structure constants of U(N) are:
\begin{eqnarray}\label{A1}
\hspace{-0.5cm} C^{abd}C^{dcm}+C^{cad}C^{dbm}+C^{bcd}C^{dam}=0,&\hspace{0cm}&
D^{abd}D^{dcm}-D^{cad}D^{dbm}-C^{bcd}C^{dam}=0,\nonumber\\
\hspace{-0.5cm}D^{abd}D^{dcm}+C^{cad}C^{dbm}-D^{bcd}D^{dam}=0,&\hspace{0.cm}&
C^{abd}C^{dcm}-D^{cad}D^{dbm}+D^{bcd}D^{dam}=0,\nonumber\\
\hspace{-0.5cm}D^{abd}C^{dcm}+D^{cad}C^{dbm}+D^{bcd}C^{dam}=0,&\hspace{0cm}&
C^{abd}D^{dcm}-C^{cad}D^{dbm}+D^{bcd}C^{dam}=0,\nonumber\\
\hspace{-0.5cm}C^{abd}D^{dcm}-D^{cad}C^{dbm}-C^{bcd}D^{dam}=0,&\hspace{0cm}&
D^{abd}C^{dcm}+C^{cad}D^{dbm}-C^{bcd}D^{dam}=0,
\end{eqnarray} 
\par\vskip0.5cm\noindent
{\it Nonplanar Integrals}
\par\vskip0.2cm\noindent
Here we want to evaluate the integrals of the generic forms:
\begin{eqnarray}\label{B1}
I_{1}&\equiv&\int\limits_{-\infty}^{+\infty}\frac{d^{D}\ell}{\left(2\pi\right)^{D}}\ \frac{\ell_{\perp}^{2}
e^{-2i\ell\times q}
}{\left(\ell^{2}+\Delta\right)^{3}},\nonumber\\
I_{\xi}&\equiv&\int\limits_{-\infty}^{+\infty}\frac{d^{D}\ell}{\left(2\pi\right)^{D}}\ \frac{\left(\ell_{\perp}\right)_{\xi}e^{-2i\ell\times q}}{\left(\ell^{2}+\Delta\right)^{3}},\nonumber\\
I_{\eta\xi}&\equiv&\int\limits_{-\infty}^{+\infty}\frac{d^{D}\ell}{\left(2\pi\right)^{D}}\ \frac{\left(\ell_{||}\right)_{\eta}\left(\ell_{\perp}\right)_{\xi}e^{-2i\ell\times q}}{\left(\ell^{2}+\Delta\right)^{3}},
\end{eqnarray}
where $q$ is an arbitrary external momentum and the integrations are over Euclidean integration variables. To evaluate the above integrals let us introduce the spherical coordinates in $D$ dimensions. The first four components are denoted by $\ell_{||}$ and the other ones by  $\ell_{\perp}$. We therefore have:
\begin{eqnarray}\label{B4}
\ell^{2}_{\perp}\equiv \ell^{2}-\sum\limits_{i=1}^{4}\ell_{i}^{2}=\ell^{2}-\ell^{2}\prod_{k=3}^{D-2}\sin^{2}\theta_{k}.
\end{eqnarray}
Using these notations the integral $I_{1}$ reads 
\begin{eqnarray}\label{B5}
I_{1}
&=&\int\limits_{0}^{\infty}\ \frac{ d^{4}\ell\ \ell^{D-2}e^{-2i\ell\times q}}{\left(\ell^{2}+\Delta\right)^{3}}\int\limits_{0}^{\pi}\left(\prod\limits_{k=3}^{D-2}\sin^{k}\theta_{k}d\theta_{k}-\prod\limits_{k=3}^{D-2}\sin^{k+2}\theta_{k}d\theta_{k}\right), 
\end{eqnarray}
which turns out to be 
\begin{eqnarray}\label{B8}
I_{1}=\frac{\pi^{\frac{D}{2}-2}}{\Gamma\left(D/2\right)}\frac{D-4}{D}\int\limits_{0}^{\infty}\ \frac{d^{4}\ell\ \ell^{2}e^{-2i\ell\times q}}{\left(\ell^{2}+\Delta\right)^{3}}.
\end{eqnarray}
\par\noindent $\bullet$ For $I_{\xi}$ and $I_{\eta\xi}$ use 
\begin{eqnarray}\label{B9}
\left(\ell_{\perp}\right)_{\xi}\equiv \ell\prod\limits_{k=\xi-1}^{D-2}\sin\theta_{k}\cos\theta_{\xi-2},\hspace{0.5cm}\mbox{for}\hspace{0.5cm}5\leq\xi\leq D, 
\end{eqnarray} 
and get  
\begin{eqnarray}\label{B10}
\lefteqn{\hspace{-1cm}
I_{\xi}=
\int\frac
{d^{4}\ell\ell^{D-3}e^{-2i\ell\times q}}
{\left(\ell^{2}+\Delta\right)^{3}} 
\prod\limits_{m=\alpha-1}^{D-2}
\sin\theta_{m}\cos\theta_{\alpha-2}
\prod\limits_{k=3}^{D-2}
\sin^{k}\theta_{k}d\theta_{k} }\nonumber\\
&&\hspace{-1.3cm}=\int\frac{d^{4}\ell\ell^{D-3}e^{-2i\ell\times q}}{\left(\ell^{2}+\Delta\right)^{3}}
\int\limits_{0}^{\pi}\prod\limits_{k=3}^{\alpha-3}\sin^{k}\theta_{k}
\int\limits_{0}^{\pi}\sin^{\alpha-2}\theta_{\alpha-2}\cos\theta_{\alpha-2}d\theta_{\alpha-2}
\int\limits_{0}^{1}\prod\limits_{k=\alpha-1}^{D-2}\sin^{k+1}\theta_{k}d\theta_{k}.
\end{eqnarray}
Using
$\int\limits_{0}^{\pi}d\theta_{\alpha-2}\sin^{\alpha-2}\theta_{\alpha-2}\cos\theta_{\alpha-2}=0,$ for $\alpha=5,\cdots,D$ we have $I_{\xi}=0$. It is simply seen that $I_{\eta\xi}=0$
for $1\leq\eta\leq 4$ and $5\leq \xi\leq D$. 
\par\noindent $\bullet$
For 
\begin{eqnarray}\label{B12}
I_{2}\equiv
\int\limits_{-\infty}^{+\infty}\frac{d^{D}\ell}{\left(2\pi\right)^{D}}\ 
\frac{\ell^{2}e^{-2i\ell\times q}}
{\left(\ell^{2}+\Delta\right)^{3}},
\end{eqnarray}
using the parametrization
\begin{eqnarray}\label{B13}
\frac{1}{\ell^{2}+\Delta}=\int\limits_{0}^{\infty}e^{-\alpha\left(\ell^{2}+\Delta\right)}d\alpha,
\end{eqnarray}
we have
\begin{eqnarray}\label{B14}
I_{2}=\sum\limits_{\xi}\frac{\partial^{2}}{\partial z_{\xi}^{2}}
\int\limits_{-\infty}^{+\infty}\frac{d^{D}\ell}{\left(2\pi\right)^{D}}\ 
\int\limits_{0}^{\infty}\mbox{exp}\bigg[z_{\eta}\ell_{\eta}-\ell^{2}\left(\alpha_{1}+\alpha_{2}+\alpha_{3}\right)-\Delta\left(\alpha_{1}+\alpha_{2}+\alpha_{3}\right)-i\tilde{q}_{\eta}\ell_{\eta}\bigg]d\alpha_{1}d\alpha_{2}d\alpha_{3},\nonumber\\
\end{eqnarray}
where $\tilde{q}_{\eta}\equiv \theta_{\eta\sigma}q_{\sigma}$. Performing the $\ell$-integration we obtain
\begin{eqnarray}\label{B16}
\lefteqn{\hspace{-1cm}
I_{2}=\frac{D}{2\left(2\pi\right)^{D/2}}\int\limits_{0}^{\infty}\frac{d\alpha_{1}d\alpha_{2}d\alpha_{3}}{\left(\alpha_{1}+\alpha_{2}+\alpha_{3}\right)^{D/2+1}}\ \mbox{exp}\bigg[
\frac{\left(i\tilde{q}_{\eta}\right)^{2}}{4\left(\alpha_{1}+\alpha_{2}+\alpha_{3}\right)}-\Delta\left(\alpha_{1}+\alpha_{2}+\alpha_{3}\right)
\bigg]
}\nonumber\\
&&\hspace{-1cm}+\frac{\left(i\tilde{q}_{\eta}\right)^{2}}{4\left(4\pi\right)^{D/2}}\int\limits_{0}^{\infty}\frac{d\alpha_{1}d\alpha_{2}d\alpha_{3}}{\left(\alpha_{1}+\alpha_{2}+\alpha_{3}\right)^{D/2+2}}\ \mbox{exp}\bigg[
\frac{\left(i\tilde{q}_{\eta}\right)^{2}}{4\left(\alpha_{1}+\alpha_{2}+\alpha_{3}\right)}-\Delta\left(\alpha_{1}+\alpha_{2}+\alpha_{3}\right)\bigg].
\end{eqnarray}
Inserting  
\begin{eqnarray}\label{B17}
1=\int\limits_{0}^{\infty}d\rho\ \delta\left(\rho-\sum\limits_{\ell=1}^{3}\alpha_{\ell}\right),
\end{eqnarray}
on the r.h.s. of (\ref{B16}) and rescaling the Feynman parameters $\alpha_{\ell}\to \rho\alpha_{\ell}$ for $\ell=1,2,3$, we arrive at:
\begin{eqnarray}\label{B18}
I_{2}=\frac{D}{2\left(4\pi\right)^{D/2}}\int\limits_{0}^{\infty}\frac{d\rho}{\rho^{D/2-1}}\ \mbox{exp}\bigg[-\frac{q\circ q}{4\rho}-\Delta\rho\bigg]-
\frac{\tilde{q}^{2}}{4\left(4\pi\right)^{D/2}}\int\limits_{0}^{\infty}
\frac{d\rho}{\rho^{D/2}}\ \mbox{exp}\bigg[-\frac{q\circ q}{4\rho}-\Delta\rho\bigg],
\end{eqnarray}  
where $\tilde{q}^{2}\equiv q\circ q\equiv -q_{\sigma}\theta_{\sigma\eta}\theta_{\eta\xi}\ q_{\xi}$.
This is the result used in Eqs. (\ref{M3-2-10a})-(\ref{M3-2-10b}).
\end{appendix}
\vskip4cm

\begin{table}[ht]{\large\bf Table 1}\par \vskip0.3cm
\begin{tabular}{|lcl||ccl|}\hline
\multicolumn{3}{|c||}{Phases}&
\multicolumn{3}{|c|}{Group factors}\\ \hline\hline
$f_{A}^{1}$&=&$e^{ik_{2}\times k_{3}}$&  
$\big[{\cal{G}}_{A}^{cfm}]_{1}$&\hspace{-0.2cm}=& 
$\hspace{-0.3cm}+K^{abc}K^{daf}K^{bdm}-K^{abc}K^{daf}iC^{bdm}$\\
&&&&&$\hspace{-0.3cm}-K^{abc}iC^{daf}K^{bdm} -iC^{abc}K^{daf}K^{bdm}$\\
&&&&&$\hspace{-0.3cm}+K^{abc}iC^{daf}iC^{bdm}+iC^{abc}K^{daf}iC^{bdm}$\\
&&&&&$\hspace{-0.3cm}+iC^{abc}iC^{daf}K^{bdm}-iC^{abc}iC^{daf}iC^{bdm}$\\
$f_{A}^{2}\left(\ell\right)$&=&$e^{-ik_{2}\times k_{3}-2i\ell\times k_{3}}$&$ \big[{\cal{G}}_{A}^{cfm}]_{2} $&\hspace{-.2cm}=&$\hspace{-0.3cm}+K^{abc}K^{daf}K^{bdm}-K^{abc}K^{daf}iC^{bdm}$    \\
$f_{A}^{3}\left(\ell\right)$&=&$e^{-ik_{2}\times k_{3}-2i\ell\times \left(k_{2}+k_{3}\right)}$ &$\big[{\cal{G}}_{A}^{cfm}]_{3}$&\hspace{-0.2cm}=&$\hspace{-0.3cm}-K^{abc}K^{daf}K^{bdm}+K^{abc}K^{daf}iC^{bdm} $\\
&&&&&$\hspace{-0.3cm}+K^{abc}iC^{daf}K^{bdm} -K^{abc}iC^{daf}iC^{bdm}$  \\
$f_{A}^{4}\left(\ell\right)$&=&$e^{-ik_{2}\times k_{3}-2i\ell\times  k_{2}} $ &$\big[{\cal{G}}_{A}^{cfm}]_{4}$&\hspace{-0.2cm}=&$\hspace{-0.3cm}+K^{abc}K^{daf}K^{bdm}-K^{abc}iC^{daf}K^{bdm}$      \\
$f_{A}^{5}\left(\ell\right)$&=&$e^{+ik_{2}\times k_{3}+2i\ell\times  k_{2}}$&$\big[{\cal{G}}_{A}^{cfm}]_{5}$&\hspace{-0.2cm}=&
$\hspace{-0.3cm}-K^{abc}K^{daf}K^{bdm}+K^{abc}K^{daf}iC^{bdm}  $\\
&&&&&$\hspace{-0.3cm} +iC^{abc}K^{daf}K^{bdm}-iC^{abc}K^{daf}iC^{bdm}$ \\
$f_{A}^{6}\left(\ell\right)$&=&$e^{+ik_{2}\times k_{3}+2i\ell\times k_{3}}$&$\big[{\cal{G}}_{A}^{cfm}]_{6}$&\hspace{-0.2cm}=&
$\hspace{-0.3cm}-K^{abc}K^{daf}K^{bdm}+K^{abc}iC^{daf}K^{bdm}  $\\
&&&&&$\hspace{-0.3cm}  +iC^{abc}K^{daf}K^{bdm}-iC^{abc}iC^{daf}K^{bdm}$\\
$f_{A}^{7}\left(\ell\right)$&=&$e^{+ik_{2}\times k_{3}+2i\ell\times \left(k_{2}+k_{3}\right)}$&$\big[{\cal{G}}_{A}^{cfm}]_{7}$
&\hspace{-0.2cm}=&$\hspace{-0.3cm}+K^{abc}K^{daf}K^{bdm}-iC^{abc}K^{daf}K^{bdm}$     \\  
$f_{A}^{8}$&=&$e^{-ik_{2}\times k_{3}}$&$\big[{\cal{G}}_{A}^{cfm}]_{8}$&\hspace{-0.2cm}=&$\hspace{-0.3cm}-K^{abc}K^{daf}K^{bdm}$    \\
\hline
\end{tabular}
\end{table}
\newpage
\begin{table}[ht]{\large\bf Table 2}\par \vskip0.3cm
\begin{tabular}{|lcl||ccl|}\hline
\multicolumn{3}{|c||}{Phases}&
\multicolumn{3}{|c|}{Group factors}\\ \hline\hline
$f_{B}^{1}$&=&$e^{ik_{2}\times k_{3}}$&$ \big[{\cal{G}}_{B}^{cfm}]_{1} $&\hspace{-0.2cm}=&$\hspace{-0.3cm} -K^{abc}K^{bdf}K^{dam}$    \\
$f_{B}^{2}\left(\ell\right)$&=&$e^{-ik_{2}\times k_{3}+2i\ell\times k_{3}} $&$\big[{\cal{G}}_{B}^{cfm}]_{2}$&\hspace{-0.2cm}=&$\hspace{-0.3cm}
-K^{abc}K^{bdf}K^{dam}+K^{abc}iC^{bdf}K^{dam}  $\\
&&&&&$\hspace{-0.3cm}+iC^{abc}K^{bdf}K^{dam}-iC^{abc}iC^{bdf}K^{dam}$    \\
$f_{B}^{3}\left(\ell\right)$&=&$e^{-ik_{2}\times k_{3}+2i\ell\times \left(k_{2}+k_{3}\right)}$&$\big[{\cal{G}}_{B}^{cfm}]_{3}$&\hspace{-0.2cm}=&$\hspace{-0.3cm} +K^{abc}K^{bdf}K^{dam} -iC^{abc}K^{bdf}K^{dam}$    \\
$f_{B}^{4}\left(\ell\right)$&=&$e^{-ik_{2}\times k_{3}+2i\ell\times  k_{2}}$&$\big[{\cal{G}}_{B}^{cfm}]_{4}$&\hspace{-0.2cm}=&$\hspace{-0.3cm}
-K^{abc}K^{bdf}K^{dam}+K^{abc}K^{bdf}iC^{dam}  $\\
&&&&&$\hspace{-0.3cm}  +iC^{abc}K^{bdf}K^{dam}-iC^{abc}K^{bdf}iC^{dam}$\\
$f_{B}^{5}\left(\ell\right)$&=&$e^{+ik_{2}\times k_{3}-2i\ell\times  k_{2}}$&$\big[{\cal{G}}_{B}^{cfm}]_{5}$&\hspace{-0.2cm}=&$\hspace{-0.3cm}+K^{abc}K^{bdf}K^{dam}-K^{abc}iC^{bdf}K^{dam}$   \\
$f_{B}^{6}\left(\ell\right)$&=&$e^{+ik_{2}\times k_{3}-2i\ell\times k_{3}}$&$\big[{\cal{G}}_{B}^{cfm}]_{6}$&\hspace{-0.2cm}=&$\hspace{-0.3cm}
+K^{abc}K^{bdf}K^{dam}-K^{abc}K^{bdf}iC^{dam}  $   \\
$f_{B}^{7}\left(\ell\right)$&=&$e^{+ik_{2}\times k_{3}-2i\ell\times \left(k_{2}+k_{3}\right)}$&$\big[{\cal{G}}_{B}^{cfm}]_{7}$&\hspace{-0.2cm}=&  $\hspace{-0.3cm}
-K^{abc}K^{bdf}K^{dam}+K^{abc}K^{bdf}iC^{dam}  $\\
&&&&&$ \hspace{-0.3cm}+K^{abc}iC^{bdf}K^{dam}-K^{abc}iC^{bdf}iC^{dam}$  \\ 
$f_{B}^{8}$&=&$e^{-ik_{2}\times k_{3}}$&$\big[{\cal{G}}_{B}^{cfm}]_{8}$&\hspace{-0.2cm}=& $\hspace{-0.3cm}
+K^{abc}K^{bdf}K^{dam}-K^{abc}K^{bdf}iC^{dam}  $\\
&&&&&$\hspace{-0.3cm}-K^{abc}iC^{bdf}K^{dam}-iC^{abc}K^{bdf}K^{dam}$\\
&&&&&$\hspace{-0.3cm}+K^{abc}iC^{bdf}iC^{dam}+iC^{abc}K^{bdf}iC^{dam}$\\
&&&&&$\hspace{-0.3cm}+iC^{abc}iC^{bdf}K^{dam}-iC^{abc}iC^{bdf}iC^{dam}$\\
\hline      
\end{tabular}
\end{table}

\newpage
\section{Figures}
\vskip2cm

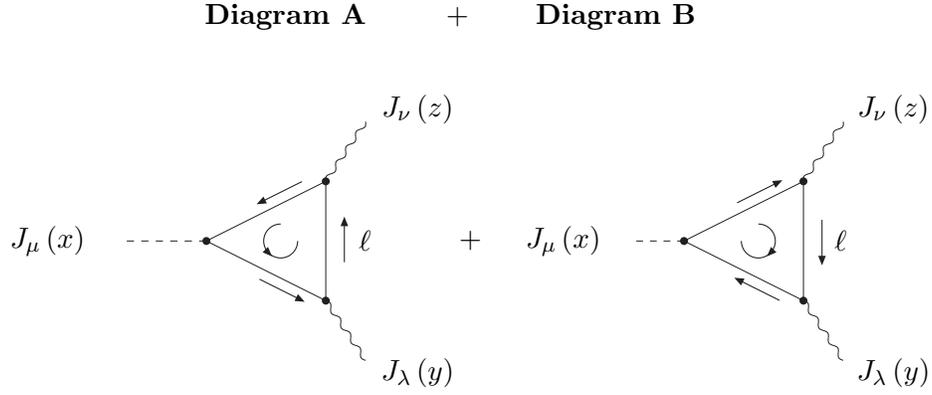
\begin{figure}[h]                              
\SetScale{0.3}
\begin{center}
\begin{picture} (200,40)(0,20)
\Vertex(0,0){5}
\Vertex(150,75){5}
\Vertex(150,-75){5}
\Line(0,0)(150,75)
\Line(0,0)(150,-75)
\Line(150,75)(150,-75)
\DashLine(-100,0)(0,0){10}
\Photon(150,75)(200,150){3}{4}
\Photon(150,-75)(200,-150){3}{4}
\Text(60,0)[]{$\ell$}
\Text(-60,0)[]{$J_{\mu}\left(x\right)$}
\Text(80,50)[]{$J_{\nu}\left(z\right)$}
\Text(80,-50)[]{$J_{\lambda}\left(y\right)$}
\SetScale{0.8}
\LongArrow(65,-10)(65,10)
\LongArrow(45,28)(25,18)
\LongArrow(25,-18)(45,-28)
\ArrowArc(35,0)(8,90,360)
\SetScale{0.3}
\Text(30,85)[]{\bf Diagram A}
\Text(95,85)[]{$+$}
\Text(155,85)[]{\bf Diagram B}
\Text(100,0)[]{$+$}
\Vertex(600,0){5}
\Vertex(750,75){5}
\Vertex(750,-75){5}
\Line(600,0)(750,75)
\Line(600,0)(750,-75)
\Line(750,75)(750,-75)
\DashLine(540,0)(600,0){10}
\Photon(750,75)(800,150){3}{4}
\Photon(750,-75)(800,-150){3}{4}
\Text(240,0)[]{$\ell$}
\Text(135,0)[]{$J_{\mu}\left(x\right)$}
\Text(260,50)[]{$J_{\nu}\left(z\right)$}
\Text(260,-50)[]{$J_{\lambda}\left(y\right)$}
\SetScale{0.8}
\LongArrow(290,10)(290,-10)
\LongArrow(250,18)(270,28)
\LongArrow (270,-28)(250,-18)
\ArrowArcn(260,0)(8,90,180)
\end{picture}
\end{center}
\vskip2.5cm
\caption{Triangle Diagrams for the gauge anomaly in the current $J_{\mu}\left(x\right)$ indicated by the dashed line. Each diagram is decorated by a set of planar and nonplanar phases. For U(N) chiral gauge theory the currents are to be replaced by $J_{\mu}^{c}\left(x\right), J_{\lambda}^{f}\left(y\right), $ and $J_{\nu}^{m}\left(z\right)$. For non-chiral U(1) gauge theory only $J_{\mu}\left(x\right)$ is to be replaced by the anomalous $J_{\mu,5}\left(x\right)$.}\label{Fig1}
\end{figure}
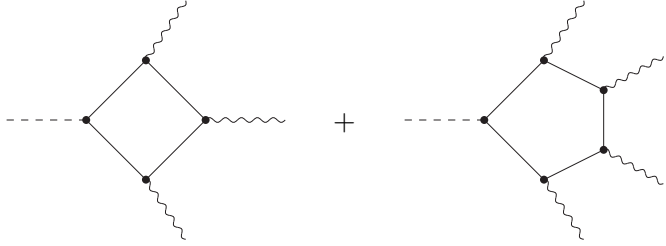
\begin{figure}[htbp]                              
\SetScale{0.3}
\begin{center}
\begin{picture} (350,60)(0,0)

\Vertex(200,20){5}
\Vertex(350,20){5}
\Vertex(275,95){5}                                                                         
\Vertex(275,-55){5}
\DashLine(100,20)(200,20){10}
\Line(200,20)(275,95)
\Line(200,20)(275,-55)
\Line(350,20)(275,95)
\Line(350,20)(275,-55)
\Photon(275,95)(325,170){3}{5}
\Photon(275,-55)(325,-130){3}{5}
\Photon(350,20)(450,20){3}{5}
\Text(158,5)[]{$+$}
\Vertex(700,20){5}
\Vertex(775,95){5}
\Vertex(775,-55){5}
\Vertex(850,57.5){5}
\Vertex(850,-17.5){5}
\DashLine(600,20)(700,20){10}
\Line(700,20)(775,95)
\Line(700,20)(775,-55)
\Line(850,57.5)(775,95)
\Line(850,-17.5)(775,-55)
\Line(850,57.5)(850,-17.5)
\Photon(775,95)(825,170){3}{5}
\Photon(775,-55)(825,-130){3}{5}
\Photon(850,57.5)(925,100){3}{5}
\Photon(850,-17.5)(925,-60){3}{5}
\end{picture}
\end{center}
\vskip2cm
\caption{Higher loop diagrams contributing to anomaly.}\label{Fig2}
\end{figure}

\end{document}